\documentclass[floatfix, aps, prb, superscriptaddress, reprint]{revtex4-1}

\usepackage{graphicx}               
\usepackage{bm}                     
\usepackage{amsfonts,amssymb}       
\usepackage{dcolumn}                
\usepackage{multirow}

\newcommand*\pct{\scalebox{.9}{\%}}

\begin{document}

\title{Diverse anisotropy of phonon transport in
       two-dimensional group IV-VI compounds:\\
       A comparative study}

\author{Guangzhao~Qin}
\affiliation{College of Materials Science and Opto-Electronic Technology, University of Chinese Academy of Sciences, Beijing 100049, China}
\affiliation{Institute of Mineral Engineering, Division of Materials Science and Engineering, Faculty of Georesources and Materials Engineering, RWTH Aachen University, Aachen 52064, Germany}
\author{Zhenzhen~Qin}
\affiliation{College of Electronic Information and Optical Engineering, Nankai University, Tianjin 300071, China}
\author{Wu-Zhang Fang}
\affiliation{School of Physics, University of Chinese Academy of Sciences, Beijing 100049, China}
\author{Li-Chuan Zhang}
\affiliation{College of Materials Science and Opto-Electronic Technology, University of Chinese Academy of Sciences, Beijing 100049, China}
\author{Sheng-Ying~Yue}
\affiliation{Aachen Institute for Advanced Study in Computational Engineering Science (AICES), RWTH Aachen University, Aachen 52062, Germany}
\author{Qing-Bo~Yan}
\email{yan@ucas.ac.cn}
\affiliation{College of Materials Science and Opto-Electronic Technology, University of Chinese Academy of Sciences, Beijing 100049, China}
\author{Ming~Hu}
\email{hum@ghi.rwth-aachen.de}
\affiliation{Institute of Mineral Engineering, Division of Materials Science and Engineering, Faculty of Georesources and Materials Engineering, RWTH Aachen University, Aachen 52064, Germany}
\affiliation{Aachen Institute for Advanced Study in Computational Engineering Science (AICES), RWTH Aachen University, Aachen 52062, Germany}
\author{Gang~Su}
\email{gsu@ucas.ac.cn}
\affiliation{School of Physics, University of Chinese Academy of Sciences, Beijing 100049, China}

\date{\today}

\begin{abstract}
New classes two-dimensional (2D) materials beyond graphene, including layered and
non-layered, and their heterostructures, are currently attracting increasing
interest due to their promising applications in nanoelectronics, optoelectronics
and clean energy, where thermal transport property is one of the fundamental
physical parameters.
In this paper, we systematically investigated the phonon transport properties of
2D orthorhombic group IV-VI compounds of $GeS$, $GeSe$, $SnS$ and $SnSe$ by
solving the Boltzmann transport equation (BTE) based on first-principles
calculations.
Despite the similar puckered (hinge-like) structure along the armchair direction
as phosphorene, the four monolayer compounds possess diverse anisotropic
properties in many aspects, such as phonon group velocity, Young's modulus and
lattice thermal conductivity ($\kappa$), etc.
Especially, the $\kappa$ along the zigzag and armchair directions of monolayer
$GeS$ shows the strongest anisotropy while monolayer $SnS$ and $SnSe$ shows an
almost isotropy in phonon transport.
The origin of the diverse anisotropy is fully studied and the underlying
mechanism is discussed in detail.
With limited size, the $\kappa$ could be effectively lowered, and the anisotropy
could be effectively modulated by nanostructuring, which would extend the
applications in nanoscale thermoelectrics and thermal management.
Our study offers fundamental understanding of the anisotropic phonon transport
properties of 2D materials, and would be of significance for further study,
modulation and applications in emerging technologies.
\end{abstract}

\maketitle

\section{Introduction}              

The bulk orthorhombic group IV-VI compounds of $GeS$, $GeSe$, $SnS$ and $SnSe$
possessing puckered (hinge-like) layered structure similar to black phosphorous
have become a hot spot of recent researches.
\cite{Nature.2014.508.7496.373-377, Science.2016.351.6269.141-144,
PhysRevB.41.5227, PhysRevB.92.115202, apl.10510101907.1.4895770, jap.1.4907805}
The advantages of group IV-VI compounds, such as earth-abundance, environmental
compatibility, less toxicity, and chemical stability, make them very attractive
for large-scale applications in photovoltaics and thermoelectrics.
\cite{Science.2016.351.6269.141-144, Nature.2014.508.7496.373-377, C4TA01643B,
C4TA04462B, apl.1.4866861, jap.1.4907805}
There have been a lot of studies on their optical properties including the
photoconductivity, refractive index and infrared- and Raman-activity from
experiments
\cite{PhysRevB.15.2177, PhysRevB.90.235144, 0022-3727-18-6-003,
J.Am.Chem.Soc..2010.132.28.9519-9521}
and first-principles calculations.
\cite{PhysRevB.92.085406, AppliedPhysicsLetters.2012.100.3.032104}
The large thermopower, high power factors, and low thermal conductivities
estimated using Cahill's model\cite{PhysRevB.46.6131} make these four
orthorhombic group IV-VI compounds promising candidates for high-efficient
thermoelectric materials.
\cite{ScientificReports.2015.5..9567}
Bulk $SnSe$ is especially a robust thermoelectric candidate for energy
conversion applications in the low and moderate temperature range due to its
anisotropic and low symmetry crystal structure.
\cite{Science.2016.351.6269.141-144, Nature.2014.508.7496.373-377}

The discovery of graphene leads to an upsurge in exploring two-dimensional (2D)
materials,\cite{C4NR06523A, Nanoscale.2015.7.16.7143-7150,
Nanoscale.2016.8.1.129-135, Nanoscale.2015.7.44.18716-18724}
such as hexagonal boron nitride (\emph{h}-BN), germanene, silicene, transition
metal dichalcogenides (TMDCs) and phosphorene, which have attracted tremendous
attention due to their unique dimension-dependent properties.
\cite{BP_NATURE}
In addition to the currently available 2D materials, 2D $SnSe$ has been recently
synthesized, which is greatly expected to be potential in the applications as
photodetector, photovoltaic, piezoelectric and thermoelectric devices.
\cite{JACS.ja3108017, C1NR10084J, AppliedPhysicsLetters.2015.107.17.173104,
Nanoscale.2015.7.38.15962-15970, ding2015thermoelectric}
Monolayer $SnSe$ is also reported to be a promising 2D anisotropic semiconductor
for nanomechanics, thermoelectrics, and optoelectronics,
\cite{Sci.Rep..2016.6..19830}
and the quantum spin Hall effect was predicted in (111)-oriented thin films of
$SnSe$.
\cite{safaei2015quantum}
Beyond the specific studies on 2D $SnSe$, there have also been a lot of studies
on the optical properties of the four 2D orthorhombic group IV-VI compounds of
$GeS$, $GeSe$, $SnS$ and $SnSe$.
\cite{0022-3727-18-6-003, J.Am.Chem.Soc..2010.132.28.9519-9521,
PhysRevB.92.085406}
Furthermore, they are also reported as flexible, stable, and efficient 2D
piezoelectric materials possessing enormous, anisotropic piezoelectric effects
with the piezoelectric coefficients about two orders of magnitude larger than
those of other 2D or bulk materials, such as $MoS_2$, $GaSe$, bulk $quartz$ and
$AlN$, which are widely used in industry.
\cite{AppliedPhysicsLetters.2015.107.17.173104, arXiv1603.01791.2016}
These novel 2D anisotropic semiconductors, which attracted tremendous interest
recently, have great potential applications in nanoelectronics, optoelectronics
and thermoelectrics, calling for fundamental study of the phonon transport
properties.
However, the phonon transport properties of these 2D orthorhombic group IV-VI
compounds are still less known except monolayer $SnSe$.
\cite{Sci.Rep..2016.6..19830, Nanoscale.2015.7.38.15962-15970, ding2015thermoelectric}
A complete and comparative prediction and understanding of the underlying phonon
transport properties is the key to expand the range of their applications in
nanoelectronics, optoelectronics and thermoelectrics.

In this paper, we conduct comprehensive investigations of the diverse phonon
transport properties of 2D orthorhombic group IV-VI compounds of $GeS$,
$GeSe$, $SnS$ and $SnSe$ by solving the Boltzmann transport equation (BTE) based
on first-principles calculations.
The four monolayer compounds, although possessing similar hinge-like structure
along the armchair direction as phosphorene, show diverse anisotropic
properties in many aspects, such as phonon group velocity, Young's modulus and
lattice thermal conductivity ($\kappa$), etc.
The remainder of the paper is organized as follows.
In Sec.~\ref{methodology}, we briefly describe the methodology for the
first-principles calculations and the Boltzmann transport theory of phonon
transport.
In Sec.~\ref{structure} and \ref{dispersion}, the optimized structures and
phonon dispersions for the studied systems are presented, respectively.
In Sec.~\ref{thermal}, we study the phonon transport properties of the four
monolayer compounds with hinge-like orthorhombic structure.
Furthermore, a detailed analysis on the diverse anisotropic properties is
presented in Sec.~\ref{anisotropy} from different aspects, including mode level
$\kappa$ and average phonon group velocity, phonon scattering channels and
electron localization function (ELF).
In Sec.~\ref{size}, the effect of finite size on $\kappa$ and the anisotropy is
studied.
In Sec.~\ref{summary}, we present the summary and conclusions.

\section{Methodology}               
\label{methodology}

All the first-principles calculations are performed based on the density functional
theory (DFT) using the projector augmented wave (PAW) method \cite{PhysRevB.59.1758}
as implemented in the Vienna \emph{ab-initio} simulation package
(\texttt{\textsc{vasp}})\cite{PhysRevB.54.11169}.
The Perdew-Burke-Ernzerhof (PBE)\cite{PhysRevLett.77.3865} of generalized
gradient approximation (GGA) is chosen as the exchange-correlation functional.
The kinetic energy cutoff of wave functions is $650\,\mathrm{eV}$ for $GeS$ and
$SnS$, and $550\,\mathrm{eV}$ for $GeSe$ and $SnSe$, respectively, which are 2.5
times the maximal recommended cutoff in the pseudo-potentials.
A Monkhorst-Pack \cite{PhysRevB.13.5188} $k$-mesh of $11\times 11\times 1$ is
adopted to sample the Brillouin zone (BZ) of all the four monolayer compounds,
with the energy convergence threshold set as $10^{-8}\,\mathrm{eV}$.
In the supercell calculations for obtaining harmonic (second order) interatomic
force constants (IFCs) and anharmonic (third order) IFCs, only $\Gamma$ point in
the reciprocal space is used for the purpose of saving
resources.\cite{apl.10510101907.1.4895770}
For 2D systems, a large vacuum spacing is necessary to avoid the interactions
between layers arising from the employed periodic boundary conditions.
The vacuum spacing is set as at least $15\,\textrm{\AA}$ along the out-of-plane
direction that is sufficiently large.
Both the cell shape and volume are fully optimized and all atoms are allowed to
relax until the maximal Hellmann-Feynman force acting on each atom is no larger
than $0.001\,\textrm{eV/\AA}$.

A sufficient large supercell is necessary for the accurate prediction of phonon
dispersions, which is important for determining the phonon group velocities and
phonon-phonon interactions that is critical for predicting thermal properties.
In principle, the supercell should be constructed to ensure lattice constants to
be at least larger than $10\,\textrm{\AA}$.
To determine the supercell size used in the real-space finite displacement
difference calculations, we performed calculations with the supercell size of
$2\times 2\times 1$, $3\times 3\times 1$, $4\times 4\times 1$ and $5\times
5\times 1$.
The corresponding phonon dispersion curves show a convergent tendency as the
supercell size increasing, such as the coupling among optical and acoustic
phonon branches, the smoothness of the phonon dispersion curves, the frequency
shift of phonon modes at $\Gamma$-point, and the flat dispersions near the
$S$ symmetry point.
Since there is no difference between the results obtained with the supercell
size of $4\times 4\times 1$ and $5\times 5\times 1$, it is doubtless that the
supercell size of $5\times 5\times 1$ is a good choice.

The force constant $C_{i\alpha;j\beta}$ can be obtained from the force caused by
displacement:
\begin{equation}
C_{i\alpha;j\beta} = - \frac{F_{i\alpha}}{\Delta_{j\beta}}\ ,
\end{equation}
where $F_{i\alpha}$ is the force along the $\alpha$ direction acting on atom $i$
resulted from the displacement along the $\beta$ direction of atom $j$
($\Delta_{j\beta}$).
The displacement amplitude of atom along the $\pm x$, $\pm y$ and $\pm z$
directions are $0.01\,\textrm{\AA}$.
The space group symmetry properties are used to reduce the calculation cost and
numerical noise of the force constants, and it can also greatly simplify the
determination of the dynamical matrix that is constructed based on the force
constants.
The frequency and eigenvector forming the phonon dispersions can be obtained by
diagonalizing the dynamical matrix.

Besides the harmonic IFCs obtained above, anharmonic IFCs, which are used for
the determination of the scattering properties, are also necessary in the
calculation of lattice thermal conductivity ($\kappa$).
A $4\times 4\times 1$ supercell is constructed to get the anharmonic IFCs, and
the first-principles based real-space finite displacement difference approach is
employed.
A cutoff radius ($r_{\mathrm{cutoff}}$) is introduced to discard the
interactions between atoms with distance larger than a certain value for
practical purposes.
In principle, the $r_{\mathrm{cutoff}}$ should exceed the range of physically
relevant anharmonic interactions to get satisfactory results.
\cite{ap1.2822891, PhysRevB.86.174307}
Here, referring to the converged cutoff radius as examined in the previous work,
we choose a cutoff radius of about $6.5\,\textrm{\AA}$, which includes up to
$15^\mathrm{th}$, $14^\mathrm{th}$, $13^\mathrm{th}$ and $11^\mathrm{th}$
nearest neighbors for $GeS$, $GeSe$, $SnS$ and $SnSe$, respectively.
\cite{apl.10510101907.1.4895770}
The dielectric tensor and Born effective charges are also obtained to take
into account of long-range electrostatic interactions.
We choose the thickness of monolayer as half the length of the lattice constant
along the out-of-plane direction of the layered bulk counterpart, which contains
two layers in the conventional cell.
\cite{PhysRevB.89.155426}
The specific values of the thickness of the four monolayer compounds are listed
in Table~\ref{tab:collect}.

The lattice thermal conductivity ($\kappa$) is calculated by solving the
linearized BTE for phonons.
At thermal equilibrium, the phonons are distributed obeying the Bose-Einstein
function $f_0(\omega_{\lambda})$ in the absence of temperature gradient or other
thermodynamical forces.
\cite{Li20141747}
The $\lambda$ is the index of phonon mode comprising both phonon polarization
$p$ and wave vector $\bm{q}$.
The $\omega_{\lambda}$ is the angular frequency.
In the steady state with a temperature gradient $\nabla T$, the phonon
distribution function $f_\lambda$ deviates from $f_0(\omega_{\lambda})$ and
this deviation can be obtained from the BTE:
\begin{equation}
\label{BTE1}
\frac{\partial f_\lambda}{\partial t} = \nabla T\cdot \bm{v}_\lambda
\frac{\partial f_\lambda}{\partial T}\ ,
\end{equation}
where $\bm{v}_\lambda$ is the phonon group velocity.
The left side is the scattering term that can be determined by taking into
account the anharmonic scattering due to intrinsic phonon-phonon interactions
and the harmonic scattering due to defects and impurities such as isotopes.
\cite{Li20141747}
The right side is the diffusion term caused by the temperature gradient $\nabla T$.
Assuming a small enough $\nabla T$ in most practical situations, Eq.~(\ref{BTE1})
can be linearized in $\nabla T$ and re-written as:
\begin{equation}
\label{BTE2}
f_\lambda - f_0(\omega_{\lambda}) = -\bm{F}_\lambda \cdot \nabla T
\frac{\mathrm{d}f_0}{\mathrm{d}T}\ ,
\end{equation}
where
$\bm{F}_\lambda = \tau^0_\lambda (\bm{v}_\lambda + \bm{\Delta}_\lambda)$
when only considering the scattering mechanism of three-phonon processes.
\cite{PhysRevB.81.085205, PhysRevB.80.125203, Phys.Rev.B.1996.53.14.9064-9073,
0953-8984-20-16-165209}
Here, $\tau^0_\lambda$ is the relaxation time obtained from perturbation theory,
which is commonly used within the relaxation time approximation (RTA), and
$\bm{\Delta}_\lambda$ in the dimension of velocity is a correction of the
deviation to the RTA prediction.
The $\tau^0_\lambda$ can be computed as:
\begin{equation}
\label{tau}
\frac{1}{\tau^0_\lambda} = \frac{1}{N}
\left(
    \sum^+_{\lambda^\prime \lambda^{\prime\prime}}
     \Gamma^+_{\lambda\lambda^\prime\lambda^{\prime\prime}}
     +
        \sum^-_{\lambda^\prime \lambda^{\prime\prime}} \frac{1}{2}
        \Gamma^-_{\lambda\lambda^\prime\lambda^{\prime\prime}}
     +
        \sum_{\lambda^\prime} \Gamma_{\lambda\lambda^\prime}
\right)\ ,
\end{equation}
where $N=N_1 \times N_2 \times N_3$ is the number of discrete $\bm{q}$
sampling in the BZ, which should be tested for the convergence of $\kappa$, and
$\Gamma^+_{\lambda\lambda^\prime\lambda^{\prime\prime}}$ and
$\Gamma^-_{\lambda\lambda^\prime\lambda^{\prime\prime}}$
are three-phonon scattering rates corresponding to absorption and emission
processes of phonons, respectively,
\cite{PhysRevB.81.085205, PhysRevB.80.125203, Phys.Rev.B.1996.53.14.9064-9073,
0953-8984-20-16-165209}
and $\Gamma_{\lambda\lambda^\prime}$ is the scattering possibility resulted from
the disorder of isotopic impurity.
\cite{PhysRevB.84.125426, PhysRevB.27.858}

Eq.~(\ref{BTE2}) is numerically solved by iterating from a zero$^\mathrm{th}$
order approximation start of $\bm{F}_\lambda = \tau^0_\lambda \bm{v}_\lambda$.
If the iteration stops at the first step, the procedure is equivalent to the
operation in the RTA.
The iterative procedure has a large impact on the study of materials such as
diamond in which the normal processes play a significant role in the
phonon-phonon scattering.
\cite{PhysRevB.80.125203}
In such situations, the RTA treating the normal processes resistive cannot yield
reasonable results.
In contrast, in materials such as Si and Ge with strong Umklapp scattering,
iterating to convergence only leads to a less than $10\pct$ increase of the room
temperature $\kappa$ compared to the RTA result.
\cite{PhysRevB.81.085205}
Based on the solution, the heat flux can be calculated and the $\kappa$ can be
obtained in terms of $\bm{F}_\lambda$:
\begin{equation}
\label{kappa}
\kappa^{\alpha\beta} = \frac{1}{k_\mathrm{B} T^2 N \Omega}
    \sum_{\lambda} f_0(f_0 + 1) (\hbar\omega_\lambda)^2
    v^{\alpha}_{\lambda} F^{\beta}_{\lambda}\ \,
\end{equation}
where $k_\mathrm{B}$ is the Boltzmann constant, $\hbar$ is the Planck constant,
$\Omega$ is the volume of the unit cell, and $\alpha$ and $\beta$ denote the
Cartesian components in the conventional cubic unit cell.
The approach described above yields predictive parameter free estimate of
$\kappa$ using only basic information of the chemical structure, and has been
implemented in the \texttt{\textsc{ShengBTE}} code\cite{Li20141747,
PhysRevB.86.174307} as employed in this work.

\begin{figure}[!tb]
\centering
    \includegraphics[width=0.45\textwidth]{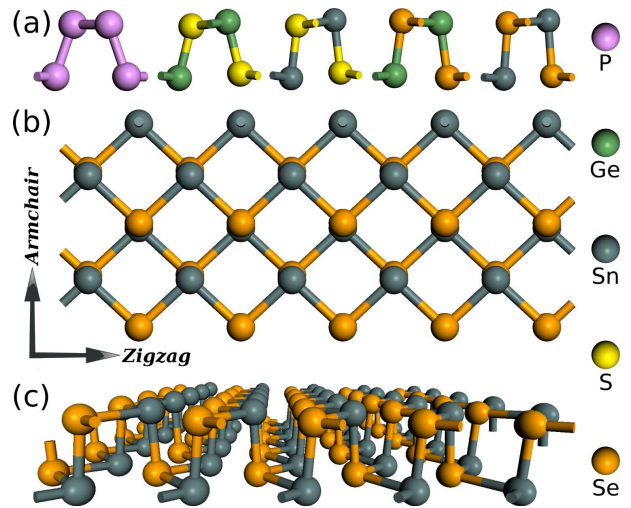}
\caption{\label{fig:structures}
(Color online)
(a) From left to right: side view of the monolayer of black phosphorus, $GeS$,
$SnS$, $GeSe$ and $SnSe$.
(b) Top view of monolayer $SnSe$. The zigzag and armchair directions are
indicated with arrows.
(c) Perspective view of monolayer $SnSe$.
The species of atoms are shown on the right.
}
\end{figure}

\begin{figure*}[!tbh]
\centering
    \includegraphics[width=0.89\textwidth]{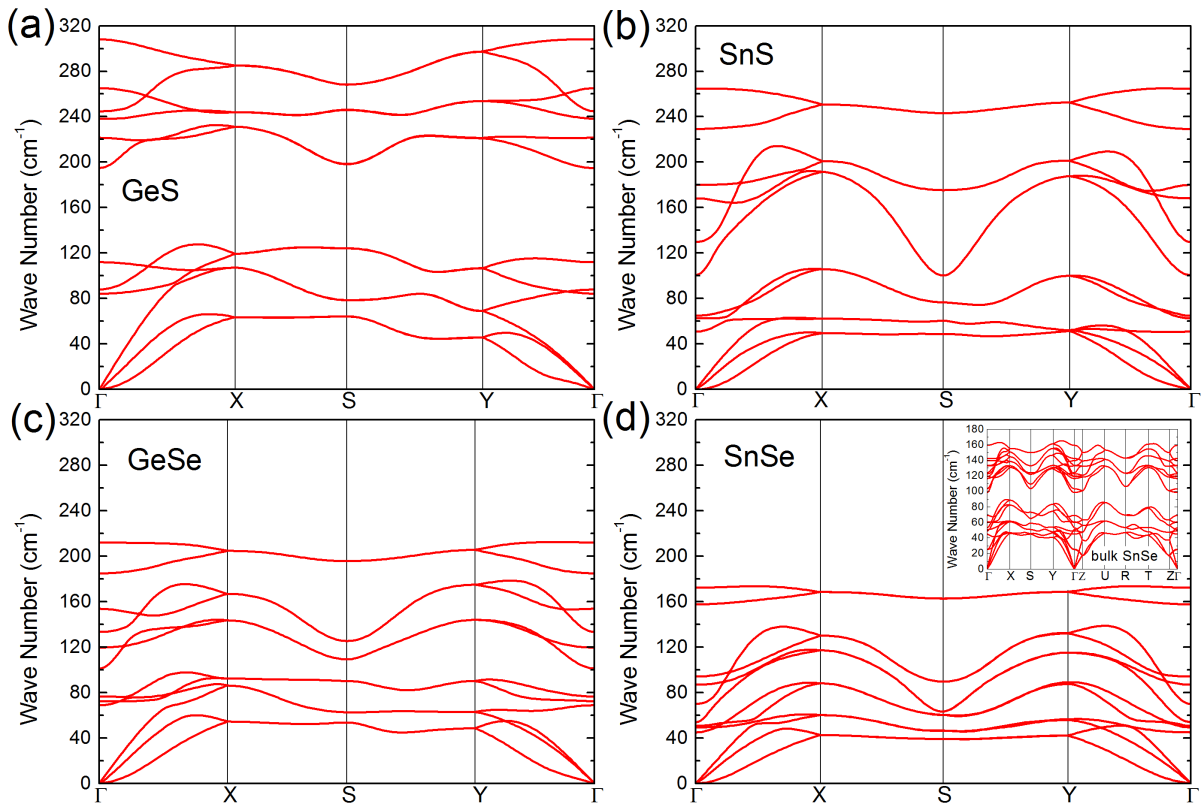}
\caption{\label{fig:dispersion}
(Color online)
The phonon dispersion curves along the path passing through the main
high-symmetry $k$-points in the irreducible Brillouin zone of monolayer (a)
$GeS$, (b) $SnS$, (c) $GeSe$ and (d) $SnSe$.
The phonon dispersion curves of bulk $SnSe$ is shown in the inset of (d) for
comparison.
The $\Gamma$-X and $\Gamma$-Y correspond to the zigzag and armchair directions,
respectively.
}
\end{figure*}

\section{Structures}                    
\label{structure}

The cell shape, volume and atoms are fully relaxed till convergence reached.
Along with the monolayer of black phosphorus that is called phosphorene, the
optimized structures of monolayer $GeS$, $SnS$, $GeSe$ and $SnSe$ are shown in
Fig.~\ref{fig:structures}(a) from left to right.
All these monolayer materials belong to the same orthorhombic crystal system,
and possess similar hinge-like structure along the armchair direction, which is
a typical feature distinctly different from the flat graphene and buckled
silicene.
As reviewed in previous work, the hinge-like structure will introduce strong
anisotropic properties.
\cite{QINsrep046946, Phys.Chem.Chem.Phys..2015..17.4854}
For example, it is harder along the zigzag direction than that along the
armchair direction, which can be verified by the Young's modulus.
The hinge-like orthorhombic structure generally leads to the lattice constants
being larger along armchair direction than along zigzag direction.
The obtained lattice constants along the two lattice directions are different,
as listed in Table~\ref{tab:collect}.
The difference from large to small is monolayer $GeS>GeSe>SnS>SnSe$.
Compared with monolayer $GeS$ and $GeSe$, the difference of lattice constants
along the two directions for monolayer $SnS$ and $SnSe$ are very small, leading
to their symmetric square-like lattice structures.
This symmetric square-like lattice structure may have significant influence on
the atomic bonding along the two directions, and will lead to their less
anisotropic properties, as we will see later.

All the space groups of the four monolayer compounds are the same
\emph{Pmn2$_1$} (No.~31), possessing 4 symmetry operators, which is different
from that of phosphorene (\emph{Pmna} (No.~53), possessing 8 symmetry
operators).
The lower symmetry of the four monolayer compounds compared with phosphorene
lies in the different types of atoms constituting the compounds, in which the
two sub-layers are not parallel to each other.
It is interesting to note from Fig.~\ref{fig:structures}(a) that the metallic
atoms $Ge$ interpose between the non-metallic (metalloid) atoms $Se$ along the
out-of-plane direction in monolayer $GeSe$, which is different from that in
monolayer $GeS$, $SnS$ and $SnSe$ where the non-metallic atoms interpose between
the metallic atoms along the out-of-plane direction.
The difference may be due to the same period of $Ge$ and $Se$ in the periodic
table resulting in the closest mass of the two atoms and the smallest
electronegativity difference in the binary compound.

\section{Phonon dispersions}            
\label{dispersion}

In the calculation of phonon dispersions, a $5\times 5\times 1$ supercell
containing 100 atoms is constructed and the first-principles based finite
displacement difference method is employed.
\cite{PhysRevB.26.3259, phonopy}
As shown in Fig.~\ref{fig:dispersion}, the phonon dispersions of monolayer
$GeS$, $SnS$, $GeSe$ and $SnSe$ have no imaginary part, indicating the
thermodynamic stability of the four monolayer compounds.
There exist 3 acoustic and 9 optical, that is 12 in total, phonon branches
corresponding to the 4 atoms per unit cell.
The three lowest phonon branches are acoustic phonon branches, i.e., the
out-of-plane flexural acoustic (FA) branch, the in-plane transverse
acoustic (TA) branch and the in-plane longitudinal acoustic (LA) branch.
The TA and LA branches is linear near the $\Gamma$ point while the FA branch is
flexural, which is similar to other 2D materials such as graphene, silicene, and
phosphorene.
\cite{Phys.Chem.Chem.Phys..2015..17.4854, PhysRevB.82.115427, RevModPhys.81.109,
PhysRevB.89.054310, AppliedPhysicsLetters.2014.104.13.131906, Phys.Rev.B.2016.93.7.075404}
This flexural feature is typically due to the 2D nature of monolayer structure.
\cite{RevModPhys.81.109}

The four monolayer compounds show very similar dispersion curves along the path
passing through the main high-symmetry $k$-points in the irreducible Brillouin
zone (IBZ) except:
1) The phonon dispersions of monolayer $GeS$ and $GeSe$ are separated into two
regions with each region possessing 6 branches.
There exists a gap between the two regions.
The gap is $2.014\,\mathrm{THz}$ for monolayer $GeS$ and $0.105\,\mathrm{THz}$
for monolayer $GeSe$.
However, there is no similar gap for monolayer $SnS$ and $SnSe$.
2) The phonon dispersions of $SnS$ and $SnSe$ are characterized by markedly
dispersive optical phonon branches, which lead to the disappearance of the gap
and result in significant group velocities of optical phonon branches.
3) The frequencies of the four monolayer compounds are different, especially
the maximum frequency ($\omega_\mathrm{M}$) of optical phonon branches at
$\Gamma$ point, and the order of $\omega_\mathrm{M}$ is monolayer
$GeS>SnS>GeSe>SnSe$.
This may be owing to the different reduced atomic mass ($\mu=m_1m_2/(m_1+m_2)$)
that the larger the $\mu$ the lower the maximum frequency.
The order of $\mu$ of the four monolayer compounds is $GeS<SnS<GeSe<SnSe$ that
is opposite to the order of $\omega_\mathrm{M}$.

From the overview of Fig.~\ref{fig:dispersion}, it is obvious that monolayer
$SnS$ and $SnSe$ possess almost symmetric phonon dispersion curves along the
$\Gamma$-X-S (left) and $\Gamma$-Y-S (right) high-symmetry $k$-paths, while the
symmetry for monolayer $GeS$ is most broken, especially for the low-frequency
phonon modes.
As a result, the anisotropy of phonon group velocity along the two different
directions of monolayer $SnS$ and $SnSe$ is tiny, and that of monolayer $GeS$
and $GeSe$ is much larger.
Based on the slope of the LA branch near the $\Gamma$ point, we could get the
group velocities along $\Gamma$-X (zigzag) and $\Gamma$-Y (armchair) directions,
respectively.
As listed in Table~\ref{tab:collect}, the specific values of the group
velocities of the four monolayer compounds show different anisotropy along the
zigzag and armchair directions.
The anisotropy of group velocities along the two directions from large to small
is monolayer $GeS>GeSe>SnS>SnSe$.
From the lattice constants as listed in Table~\ref{tab:collect}, it is found 
that the order of anisotropy of group velocities is the same as that of lattice
constants.
It is reviewed in previous work that the hinge-like structure of phosphorene is
the key to the strong anisotropy of its properties.
\cite{QINsrep046946,Phys.Chem.Chem.Phys..2015..17.4854}
However, monolayer $SnSe$ shows an almost isotropic behavior despite its similar
hinge-like structure as phosphorene.
The reason may lie in the symmetric square-like lattice structure
\cite{Sci.Rep..2016.6..19830} and the large atomic mass of $Sn$ and $Se$, which
counteract the anisotropy introduced by its hinge-like structure.

\begin{figure}[!tb]
\centering
    \includegraphics[width=0.48\textwidth]{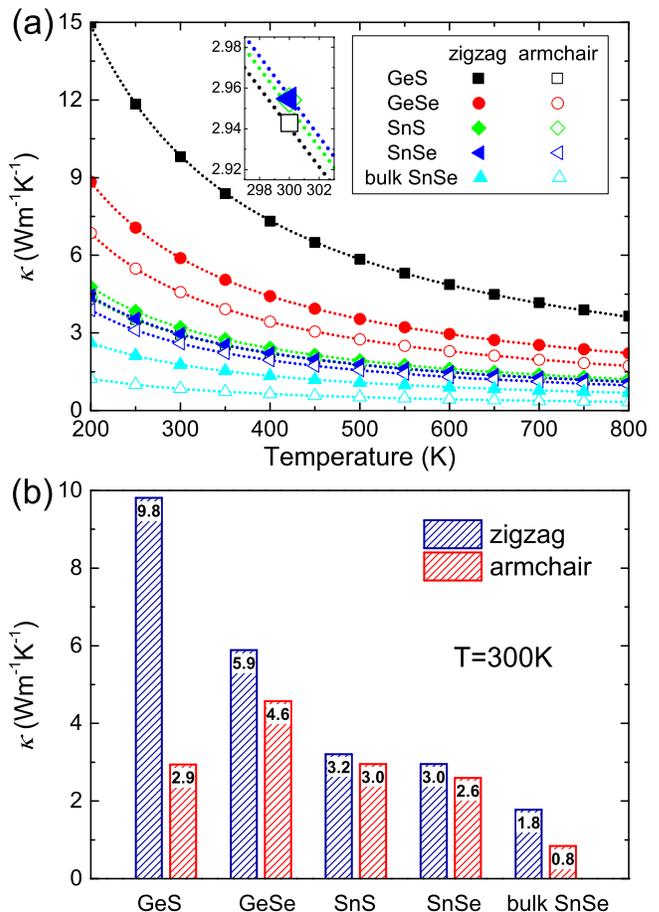}
\caption{\label{fig:LTC}
(Color online)
The lattice thermal conductivity ($\kappa$) of monolayer $GeS$, $GeSe$, $SnS$
and $SnSe$ along zigzag and armchair directions.
The in-plane $\kappa$ of bulk $SnSe$ are also plotted for comparison.
(a) $\kappa$ as a function of temperature ranging from $200\,\mathrm{K}$ to
$800\,\mathrm{K}$.
The dot lines show the fitting result of $\kappa\sim 1/T^\alpha$.
The very close $\kappa$ along the zigzag direction of monolayer $SnSe$ and along
the armchair direction of monolayer $GeS$ and $SnS$ are amplified as clearly
shown in the insert.
(b) Anisotropic $\kappa$ along zigzag and armchair directions at
$300\,\mathrm{K}$ for the four monolayer compounds and bulk $SnSe$..
}
\end{figure}

\begin{figure}[!tb]
\centering
    \includegraphics[width=0.47\textwidth]{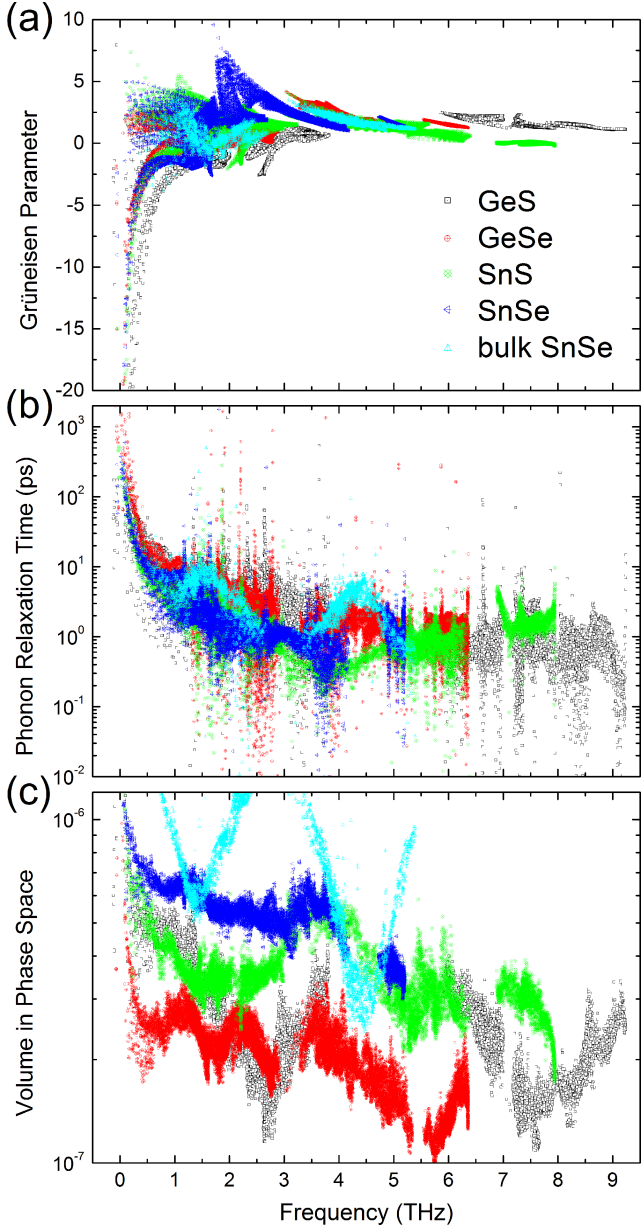}
\caption{\label{fig:last}
(Color online)
(a) Gr\"{u}neisen parameter ($\gamma$), (b) phonon relaxation time ($\tau$) and
(c) volume in phase space ($P_3$) as a function of frequency at
$300\,\mathrm{K}$ of monolayer $GeS$ (black), $GeSe$ (red), $SnS$ (green),
$SnSe$ (blue) and bulk $SnSe$ (cyan).
}
\end{figure}

\begin{figure*}[!tb]
\centering
    \includegraphics[width=0.94\textwidth]{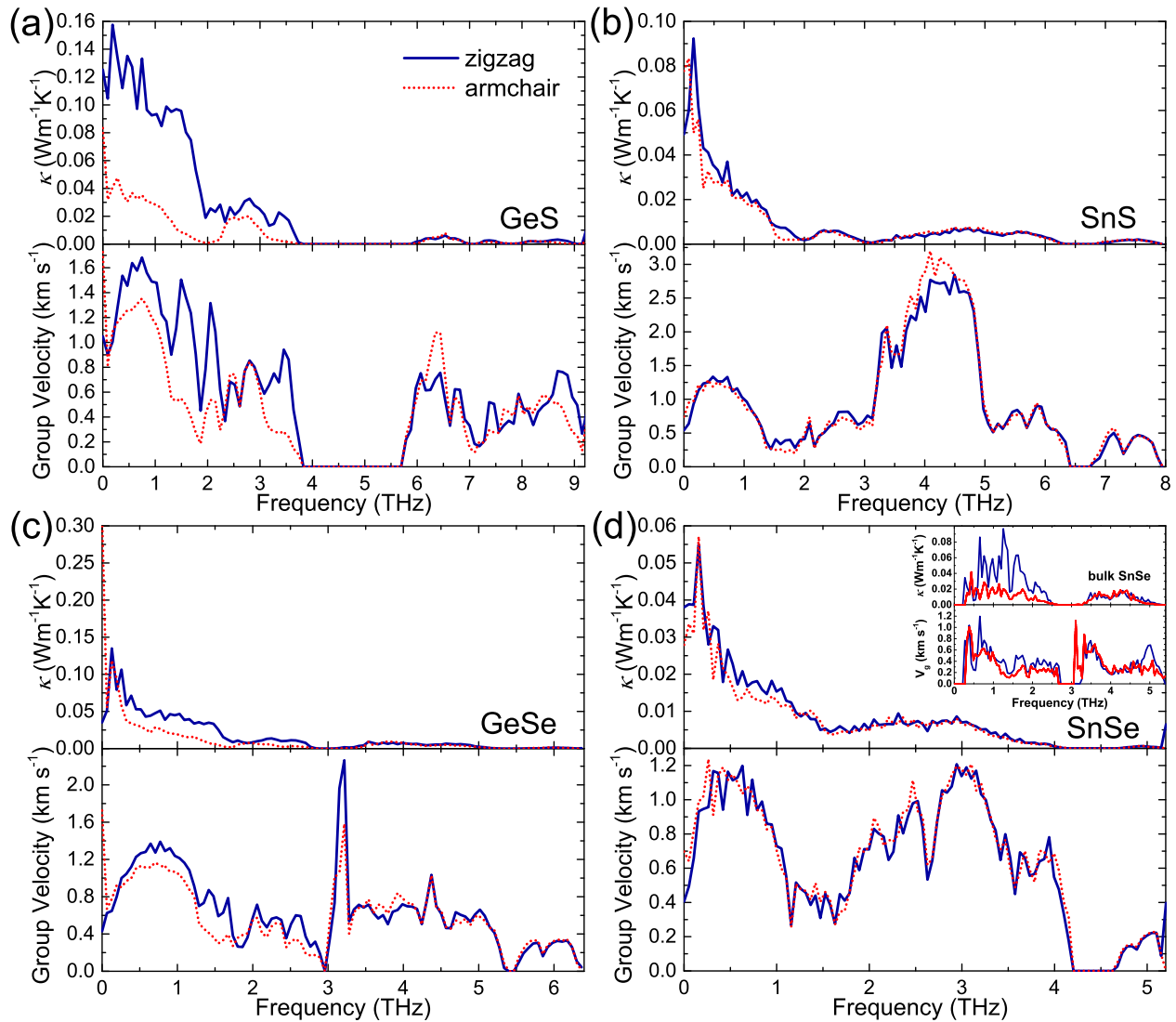}
\caption{\label{fig:kv}
(Color online)
The frequency dependent lattice thermal conductivity ($\kappa$) and average
phonon group velocity ($V_\mathrm{g}$) of monolayer (a) $GeS$, (b) $SnS$, (c)
$GeSe$ and (d) $SnSe$ along zigzag and armchair directions.
The $\kappa$ and $V_\mathrm{g}$ are summed up and averaged for all the phonon
modes in each small frequency bins, respectively.
The corresponding values of bulk $SnSe$ are shown in the inset of (d) for
comparison.
}
\end{figure*}

\begin{table*}
\small
\caption{\label{tab:collect}
Lattice constants, thickness, Young's modulus ($E$), phonon group velocity
($v_g$) at $\Gamma$ point, lattice thermal conductivity ($\kappa$) at
$300\,\mathrm{K}$, the parameter $\alpha$ in the relation of $\kappa\sim
1/T^{\alpha}$, the percentage contribution to $\kappa$ of acoustic phonon
branches (FA, TA and LA) and optical phonon branches at $300\,\mathrm{K}$ and
the representative mean free path (rMFP) at $300\,\mathrm{K}$.
The thickness of monolayer is chosen as half the length of the lattice constant
along $z$ direction of the bulk counterpart.
The properties of bulk $SnSe$ are also listed in addition to the four monolayer
compounds of $GeS$, $GeSe$, $SnS$ and $SnSe$.
}
  \begin{tabular*}{\textwidth}{@{\extracolsep{\fill}}clccccccccccc}
    \hline
    \hline
\multirow{2}{*}{Structure}  &   \multirow{2}{*}{Direction}   &         
Lattice constant & Thickness & $E$ &  $v_g$     &    $\kappa$  &
\multirow{2}{*}{$\alpha$}   & \multicolumn{4}{c}{Contribution (\%)} &
rMFP\\
\cline{9-12} & &                                                        
($\textrm{\AA}$) & ($\textrm{\AA}$) & (GPa) & (km/s) & $\mathrm{(Wm^{-1}K^{-1})}$  &
 & FA  &   TA  &   LA  & optical    &
(nm)\\
\hline                                                                  
\multirow{2}{*}{GeS} & zigzag    &
3.671   & \multirow{2}{*}{5.361} &   29.74  &
4.21  &   9.81 & 1.019 &
35.07   &   24.52   &  28.01    &   12.40   &
24.44\\
                     & armchair  &              %
4.457   &                       &   7.80   &
2.66    &   2.94   & 0.986 &
25.90   &   26.88   &   26.38   &   20.84   &
15.23\\
\multirow{2}{*}{GeSe}& zigzag    &
3.982   &   \multirow{2}{*}{5.561}  & 54.63  &
3.59  &    5.89   & 0.997 &
29.41   &   25.18   &  28.45    &   16.96   &
23.30\\
                     & armchair  &              %
4.269   &               &   22.10   &
2.73  &    4.57 & 0.996 &
37.47   &   26.47   &   20.61   &   15.45   &
72.51\\
\multirow{2}{*}{SnS} & zigzag    &
4.088   & \multirow{2}{*}{5.714} &   44.11  &
3.37  &    3.21 & 0.990 &
27.07   &   19.79   &  25.97    &   27.17   &
13.07\\
                     & armchair  &              %
4.265   &       &   25.30   &
2.94  &    2.95 & 0.984 &
34.18   &   17.67   &   18.84   &   29.31   &
13.25\\
\multirow{2}{*}{SnSe}& zigzag    &
4.294   & \multirow{2}{*}{5.888} &   45.14  &
3.13  &      2.95   & 0.999 &
31.39   &   25.16   &  14.90    &   28.55   &
12.04\\
                     & armchair  &              %
4.370   &   &   23.72  &
2.97  &    2.59 & 0.998 &
30.13   &   23.21   &   16.40   &   30.26   &
10.44\\
\multirow{2}{*}{bulk SnSe}& zigzag    &
4.214   & \multirow{2}{*}{11.776}  &   42.21  &
3.08  &      1.77   & 0.971 &
17.66   &   10.20   &  9.21    &   62.93   &
6.86\\
                     & armchair  &              %
4.520   &       &   21.16  &
2.51  &    0.84 & 0.946 &
10.96   &   13.21   &   8.86   &   66.97    &
4.30\\
    \hline
    \hline
  \end{tabular*}
\end{table*}

\section{Phonon transport properties}       
\label{thermal}

Based on the harmonic and anharmonic IFCs, the lattice thermal conductivity
($\kappa$) is calculated by solving the linearized BTE for phonons.
The phonon $Q$-grid sampling ($N\times N$) in the BZ as shown in Eq.~(\ref{tau})
has been fully tested for the convergence of $\kappa$.
The $N$ of 46, 68, 47 and 48 are chosen for getting the convergent $\kappa$ of
monolayer $GeS$, $GeSe$, $SnS$ and $SnSe$, respectively.
The much denser $Q$-grid needed for getting convergent $\kappa$ for monolayer
$GeSe$ is directly due to the much slower convergence of the $\kappa$ along
armchair direction.
The reason might lie in the fact that there are more long-wavelength phonons
contributing to the $\kappa$ along armchair direction in monolayer $GeSe$, which
would require denser $Q$-grid to capture their behavior near the zone center.
The phenomenon is consistent with the largest rMFP along armchair direction of
monolayer $GeSe$ as shown in Table~\ref{tab:collect}.
By employing the iterative method, the obtained $\kappa$ along zigzag and
armchair directions of the four monolayer compounds at different temperatures
are collected together for comparison, as shown in Fig.~\ref{fig:LTC}.
The specific values of $\kappa$ at $300\,\mathrm{K}$ are listed in
Table~\ref{tab:collect}.
We also calculated the $\kappa$ by using the RTA method, and found that 
iterating to convergence will lead to monolayer
    $GeS$: ($36.5\pct$, $22.3\pct$),
    $GeSe$: ($13.9\pct$, $8.7\pct$),
    $SnS$: ($15.8\pct$, $16.5\pct$) and
    $SnSe$: ($11.6\pct$, $11.3\pct$)
increase of the room temperature $\kappa$ compared to the RTA results, which
means stronger Umklapp scattering existing in monolayer $GeSe$, $SnS$ and $SnSe$
than monolayer $GeS$.
The two numbers in the parentheses are the ratio of increase for zigzag and
armchair directions, respectively.
The calculated $\kappa$ at different temperatures as shown in
Fig.~\ref{fig:LTC}(a) decrease with temperature increasing.
By fitting with the relation of $\kappa\sim 1/T^\alpha$, the parameter $\alpha$
are almost equal to 1, as listed in Table~\ref{tab:collect}, which is consistent
with the common behavior of $\kappa$ at medium temperatures for semi-conductors.
\cite{Phys.Chem.Chem.Phys..2015..17.4854, PhysRevB.82.035204,
PhysRevB.86.155204}

The average $\kappa$ along the two directions of the four monolayer compounds
and bulk $SnSe$ are
    $GeS$ ($6.38\,\mathrm{Wm^{-1}K^{-1}}$),
    $GeSe$ ($5.23\,\mathrm{Wm^{-1}K^{-1}}$),
    $SnS$ ($3.08\,\mathrm{Wm^{-1}K^{-1}}$),
    $SnSe$ ($2.77\,\mathrm{Wm^{-1}K^{-1}}$)
and bulk $SnSe$ ($1.31\,\mathrm{Wm^{-1}K^{-1}}$), respectively.
They possess the rather low thermal conductivity compared to a lot of materials,
such as \emph{h}-BN, MoS$_2$, graphene, silicene, phosphorene, etc.
\cite{AppliedPhysicsLetters.2013.103.25.253103, PhysRevB.89.054310,
PhysRevB.84.085204, Phys.Chem.Chem.Phys..2015..17.4854, Phys.Rev.B.2016.93.7.075404}
The $\kappa$ is strongly affected by the scattering processes, which can be
influenced by the anharmonic nature of structure and the number of allowed
three-phonon scattering processes, which are quantified as the Gr\"{u}neisen
parameters ($\gamma$) and anharmonic phase space volume ($P_3$), respectively.
\cite{apl.10510101907.1.4895770}
The order of $\kappa$ is monolayer $GeS>GeSe>SnS>SnSe>\mathrm{bulk}\ SnSe$,
which is consistent with the order of $P_3$ as shown in Fig.~\ref{fig:last}(c)
except monolayer $GeS$.
Though monolayer $GeS$ has the largest $\kappa$, it does not possess a smallest
$P_3$.
However, its $\gamma$ at low frequency range is obviously much smaller than
the other monolayer compounds.
The anharmonic nature of monolayer $GeS$ is weak compared to the other monolayer
compounds, resulting in its largest $\kappa$.
The largest $\kappa$ of monolayer $GeS$ could also be understood from its phonon
dispersion.
As compared in Sec.~\ref{dispersion}, the phonon dispersion of monolayer $GeS$
possess the largest gap of $2.014\,\mathrm{THz}$ (Fig.~\ref{fig:dispersion}(a)),
which is much larger than that of monolayer $GeSe$ ($0.105\,\mathrm{THz}$),
while monolayer $SnS$ and $SnSe$ have no gap.
The large phonon energy gap causes that the scattering of acoustic phonon modes
due to the optical phonon modes is much weaker\cite{PhysRevLett.111.025901},
which leads to the higher $\kappa$ of monolayer $GeS$.
Though there is a big gap in the phonon dispersion of bulk $SnSe$ as well, the
$\kappa$ of bulk $SnSe$ is lower than that of monolayer $SnSe$.
The reason might be that the interactions between layers in bulk $SnSe$ leads to
smaller phonon group velocity in the in-plane transport and enhanced scattering
of phonons due to the giant anharmonicity,\cite{NatPhys.2015.11.12.1063-1069}
thus result in the lower in-plane $\kappa$.
Considering the higher $\kappa$ of monolayer $SnSe$ than bulk $SnSe$, the
thermoelectric performance of monolayer $SnSe$ might be not as good as bulk
$SnSe$.
But it is hard to definitely say how the thermoelectric performance would
be, because it is determined by several strongly coupled quantities and the
thermal conductivity contributed from phonons is only one of them.
\cite{QINsrep046946}
We also notice that the $\kappa$ of bulk $SnSe$ from calculations is larger than
that from experiment.
The reason of the discrepancy may lie in two aspects:
1) The samples used for experimental measurements may have defects that would
lower the $\kappa$.
2) For such a material with giant anharmonicity,
\cite{NatPhys.2015.11.12.1063-1069}
only considering the third order IFCs for capturing the anharmonicity might be
not sufficient.
The $\kappa$ is anticipated to be further lowered if higher order IFCs are
considered.

To have a clear overview of the constituent of $\kappa$, the frequency dependent
$\kappa$ and average phonon group velocity of monolayer $GeS$, $SnS$, $GeSe$,
$SnSe$ and bulk $SnSe$ along zigzag and armchair directions are plotted in
Fig.~\ref{fig:kv}.
As shown in Fig.~\ref{fig:kv}(b) and (d), the group velocities of monolayer
$SnS$ and $SnSe$ of the optical phonon branches, especially around middle
frequency range (where the phonon dispersion can be separated into two regions
with each region containing 6 branches), are very large in a wide range.
This is consistent with the phonon dispersions of monolayer $SnS$ and $SnSe$
as shown in Fig.~\ref{fig:dispersion} that are characterized by markedly
dispersive optical phonon branches, which lead to their significant group
velocities.
From the frequency dependent $\kappa$ as shown in the top panel, it is obvious
that the $\kappa$ at middle frequency range of monolayer $SnS$ and $SnSe$
takes up a relatively larger proportion than that of monolayer $GeS$ and $GeSe$,
which is due to their relatively larger group velocities of the optical phonon
modes at middle frequency range.
The relaxation time also have some influence on the contributions to $\kappa$ of
phonon branches.
For example, as shown in Fig.~\ref{fig:last}(b), the relaxation time of
monolayer $GeS$ (black color) at high frequency range above the gap are much
smaller than that at low frequency range below the gap.
Although monolayer $GeS$ have relatively large phonon group velocity at high
frequency range, as shown in Fig.~\ref{fig:kv}(a), optical phonon branches above
the gap contribute little to $\kappa$.
As for bulk $SnSe$ (cyan color), as shown in Fig.~\ref{fig:last}(b), the
relaxation time at high frequency range above the gap of bulk $SnSe$ are
comparable to that at low frequency range below the gap, which is due to the
valley of $P_3$ as shown in Fig.~\ref{fig:last}(c).
Together with its relatively large phonon group velocity at high frequency
range, as shown in the insert in Fig.~\ref{fig:kv}(d), optical phonon branches
above the gap contribute a lot to $\kappa$.

The contributions of different phonon branches(FA, TA, LA and optical) to
$\kappa$ at $300\,\mathrm{K}$ are extracted as listed in
Table~\ref{tab:collect}.
The contributions of optical phonon branches increase with the increasing
temperature, and at the same time the contributions of acoustic phonon branches
decrease.
The cumulative $\kappa$ as a function of the phonon mean free path (MFP) at
$300\,\mathrm{K}$ for zigzag and armchair directions 
are fitted to a single parametric function
\cite{Li20141747, Phys.Chem.Chem.Phys..2015..17.4854}
\begin{equation}
\kappa(l\leq l_{max}) = \frac{\kappa_0}{1+l_0/l_{max}}\ ,
\end{equation}
respectively.
$\kappa_0$ is the ultimate cumulated $\kappa$, $l_{max}$ is the maximal MFP
concerned, and $l_0$ is the parameter to be evaluated by fitting.
The fitted curves reproduce the calculated data quite well and yield the
parameter $l_0$ for zigzag and armchair directions, respectively, which could be
interpreted as the representative MFP (rMFP).
The rMFP, as listed in Table~\ref{tab:collect}, is helpful for the
study of the size effect on the ballistic or diffusive phonon transport, which
is important for thermal design with nanostructuring.
The rMFP decrease with the increasing temperature due to the integral decrease
of MFP and higher contributions to $\kappa$ coming from optical phonon branches
at higher temperature.

Generally speaking, acoustic phonon branches possess larger MFP while optical
phonon branches possess smaller MFP, and for acoustic phonon branches lower
frequency phonon modes possess larger MFP.
Higher contributions of optical phonon branches to $\kappa$ will lead to smaller
rMFP, and higher contributions of acoustic phonon branches to $\kappa$,
especially from FA, will lead to larger rMFP.
It is noticed from Table~\ref{tab:collect} that the contributions to $\kappa$
from optical phonon branches for monolayer $SnS$ and $SnSe$ are much larger than
that for monolayer $GeS$ and $GeSe$, and bulk $SnSe$ has much larger
contributions to $\kappa$ from optical phonon branches compared with the four
monolayer compounds.
Thus bulk $SnSe$ has the smallest rMFP, and monolayer $SnS$ and $SnSe$ have
smaller rMFP than monolayer $GeS$ and $GeSe$.
Furthermore, for monolayer $GeSe$, the rMFP along zigzag direction is much
smaller than that along armchair direction, while for monolayer $SnS$, the rMFP
along zigzag direction is slightly smaller than that along armchair direction.
For other monolayer compounds, the rMFP along zigzag direction is larger than
that along armchair direction.
For monolayer $GeSe$, contributions to $\kappa$ from FA in zigzag direction is
smaller than that in armchair direction while contributions to $\kappa$ from
optical branches in zigzag direction is larger than that in armchair direction.
Both factors together result in the much smaller rMFP in zigzag direction than
that in armchair direction for monolayer $GeSe$.
For monolayer $SnS$, contributions to $\kappa$ from FA in zigzag direction is
smaller than that in armchair direction, and contributions to $\kappa$ from
optical phonon branches in zigzag direction is also smaller than that in
armchair direction.
The two competing factors result in the slightly smaller rMFP in zigzag
direction than that in armchair direction for monolayer $SnS$.
For monolayer $GeS$, $SnSe$ and bulk $SnSe$, contributions to $\kappa$ from FA
in zigzag direction is larger than that in armchair direction while
contributions to $\kappa$ from optical phonon branches in zigzag direction is
smaller than that in armchair direction, which is contrary to the case of
monolayer $GeSe$.
Both factors together result in the larger rMFP in zigzag direction than
that in armchair direction.

\section{Analysis of the anisotropy}    
\label{anisotropy}

\subsection{Anisotropy of phonon transport} 


A distinct feature in Fig.~\ref{fig:LTC} is that, all the four monolayer
compounds and bulk $SnSe$ possess anisotropic $\kappa$.
To have a clear comparison of the anisotropy, we plot the $\kappa$ along zigzag
and armchair directions at $300\,\mathrm{K}$ for the four monolayer compounds
and bulk $SnSe$ in Fig.~\ref{fig:LTC}(b).
It is obvious that the $\kappa$ of monolayer $GeS$ possesses the strongest
anisotropy compared to all the other monolayer compounds, and the $\kappa$
along the zigzag direction of monolayer $GeS$ is much larger than others.
The $\kappa$ along zigzag and armchair directions of monolayer $SnSe$ shows less
anisotropic behavior than that of bulk $SnSe$.
The order of anisotropy of the $\kappa$ of the four monolayer compounds is
monolayer $GeS>GeSe>SnSe>SnS$.
We further calculated the Young's modulus and found that its order of anisotropy
is monolayer $GeS>GeSe>SnSe>SnS$, which is identical with that of $\kappa$.
The specific values of the Young's modulus of the four monolayer compounds are
listed in Table~\ref{tab:collect}, and the related elastic constants are shown
in Table~\ref{tab:elastic}.

\begin{table}[!tbh]
\small
\caption{\label{tab:elastic}
Elastic constants with unit of kBar.
}
  \begin{tabular*}{0.5\textwidth}{@{\extracolsep{\fill}}cccccccccc}
    \hline
    \hline
\ \ \ \ \ \ \ \ \ \ \ 
            &   GeS &   GeSe    &   SnS &   SnSe    &  bulk SnSe\\
\hline
$C_{11}$    & 725.7 & 677.6     & 578.6 & 577.1     & 691.4 \\
$C_{22}$    & 204.1 & 296.0     & 272.7 & 361.7     & 355.9 \\
$C_{33}$    & -     & -         & -     & -         & 523.3 \\
$C_{44}$    & 283.2 & 319.7     & 269.2 & 286.3     & 333.0 \\
$C_{55}$    & -     & -         & -     & -         & 150.6 \\
$C_{66}$    & -     & -         & -     & -         & 118.2 \\
$C_{12}$    & 310.4 & 299.9     & 263.9 & 311.1     & 313.8 \\
$C_{13}$    & -     & -         & -     & -         & 97.6  \\
$C_{23}$    & -     & -         & -     & -         & 132.7 \\
\hline
    \hline
  \end{tabular*}
\end{table}

From Fig.~\ref{fig:kv}, it is clearly shown that the anisotropic behavior of
$\kappa$ is keeping pace with that of phonon group velocity, as
phenomenologically shown in Eq.~(\ref{kappa}).
The anisotropy of thermal conductivity is actually dominated by the anisotropy
of phonon group velocity.
\cite{AppliedPhysicsLetters.2015.106.11.111909}
A close view of Fig.~\ref{fig:kv} shows that the frequency dependent $\kappa$
and average phonon group velocity are nearly isotropic for high frequency phonon
modes above the separatrix frequency (where the gap locates), which separates
the phonon dispersion into two regions containing equal numbers of branches.
The anisotropy along the two different directions is conspicuous for low
frequency phonon modes below the separatrix frequency, which is huge for
monolayer $GeS$, small for monolayer $GeSe$, while almost disappear for
monolayer $SnS$ and $SnSe$.
Note that the phonon dispersions of monolayer $GeS$ and $GeSe$ as shown in
Fig.~\ref{fig:dispersion} are separated into two regions containing equal
numbers of branches with a gap of $2.014\,\mathrm{THz}$ and
$0.105\,\mathrm{THz}$, respectively, while there is no gap for monolayer $SnS$
and $SnSe$.
Thus we analyzed that it is the regions below the separatrix frequency where the
gap locates that result in the anisotropy.
The anisotropy is strong for monolayer $GeS$ and weak for monolayer $GeSe$ due
to the big gap in the phonon dispersions of monolayer $GeS$ and the small gap of
monolayer $GeSe$.
While there is no gap of monolayer $SnS$ and $SnSe$, their $\kappa$ and phonon
group velocity along the two different directions are almost isotropic.
Thus the anisotropy is associated with the gap of phonon dispersions that the
larger the gap the stronger the anisotropy.
If we could employ some methods such as strain or doping to modulate the gap,
the anisotropy can be effectively modulated, which would be of significance in
the thermal management applications.
The influence of the gap on the anisotropy can be understood in terms of the
coupling between the high and low frequency phonon modes at both sides of the
separatrix frequency, which is supported by the scattering channels among phonon
modes presented in the next section (Sec.~\ref{scatter}).
Since the frequency dependent average phonon group velocity of the four
monolayer compounds is nearly isotropic for high frequency phonon modes above
the separatrix frequency, the coupling will suppress the anisotropy of the low
frequency phonon modes below the separatrix frequency when there is no gap.
When a gap exists, the coupling will be weak, leading to the presence of the
anisotropy.
If the gap is very large, the coupling disappears, resulting in the appearance
of the strong anisotropy of low frequency phonon modes.


Based on the analysis demonstrated above, we also would like to present a
specific comparison for monolayer $SnSe$ and bulk $SnSe$.
Monolayer $SnSe$ shows less anisotropic properties along zigzag and armchair
directions than bulk $SnSe$.
There exist some differences between them:
1) The most significant difference is that there is no interactions between
layers in monolayer $SnSe$.
2) The difference of lattice constants along the two directions of monolayer
$SnSe$ is much less than that of bulk $SnSe$, leading to its symmetric
square-like lattice structure which determines its almost isotropic properties.
3) The Young's modulus of monolayer $SnSe$ is larger than that of bulk $SnSe$,
which means that monolayer $SnSe$ is more rigid than bulk $SnSe$, leading to the
larger phonon group velocity and $\kappa$ of monolayer $SnSe$.
4) There is no gap in the phonon dispersion of monolayer $SnSe$, while there is
a large gap for bulk $SnSe$.
The low and high frequency phonon modes at both sides of the gap are decoupled
in bulk $SnSe$ due to the large gap, resulting in the huge anisotropy of the
phonon group velocity and contributions to $\kappa$ of low frequency phonon
modes below the gap, as shown in the insert of Fig.~\ref{fig:kv}(d).
Due to the lack of the gap in the phonon dispersion of monolayer $SnSe$, the
coupling between the phonon modes will be very strong, leading to the
suppression of the anisotropy.

\begin{figure*}[!tb]
\centering
    \includegraphics[width=0.93\textwidth]{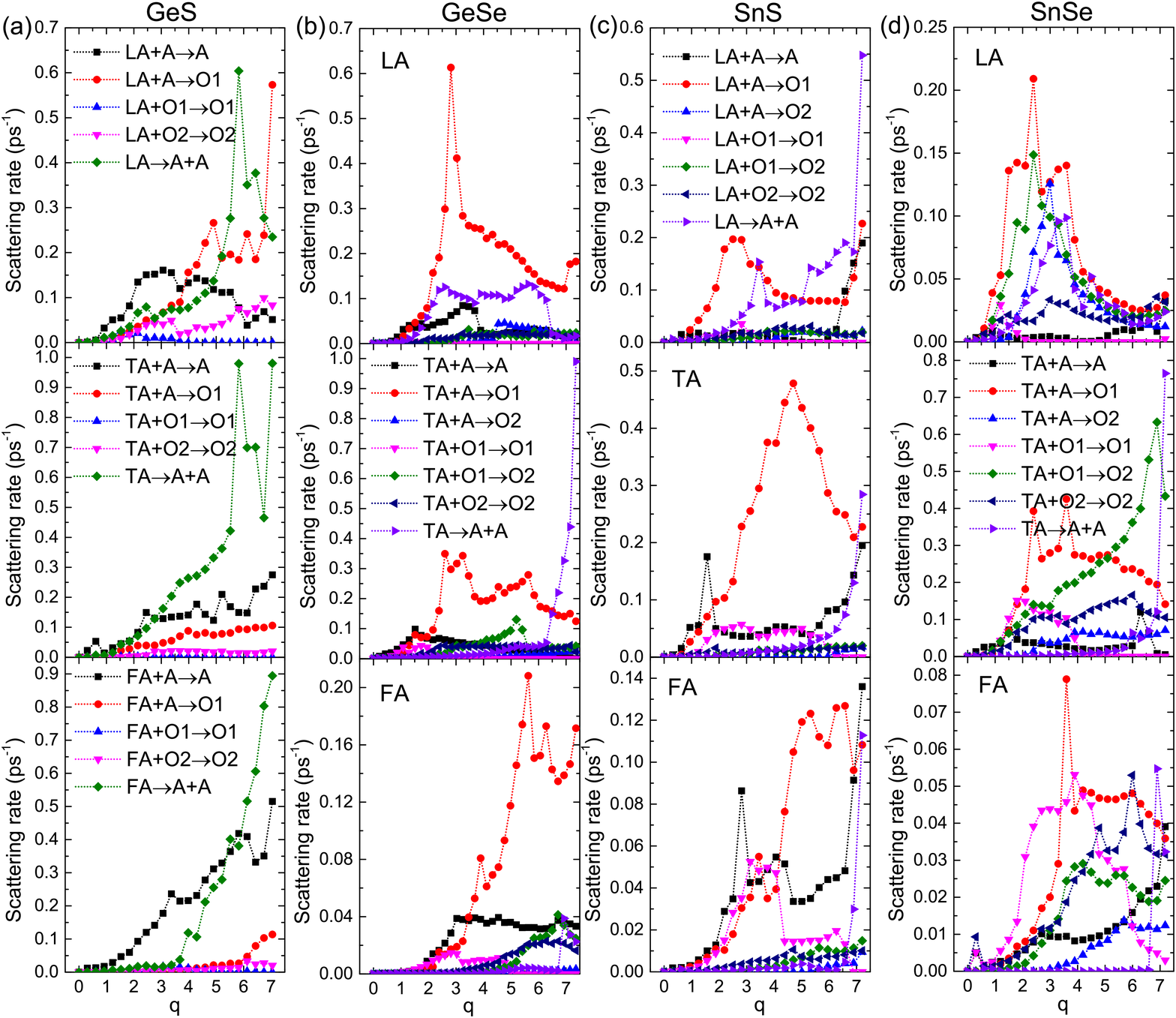}
\caption{\label{fig:scatter}
(Color online)
Scattering rate of acoustic phonon modes along $\Gamma$-Y direction for
monolayer (a) $GeS$, (b) $GeSe$, (c) $SnS$ and (d) $SnSe$ as labeled on site.
The top panel is for LA branch, the middle panel is for TA branch, and the
bottom panel is for FA branch.
In the legend, "A" means the acoustic phonon branches (LA, TA and FA), "O1"
means the 3 optical phonon branches with low frequency, "O2" means the 6 optical
phonon branches with high frequency.
The legends similar for LA, TA and FA are the same for (b) $GeSe$, (c) $SnS$ and
(d) $SnSe$, while that for (a) $GeS$ is different.
The crossover of LA, TA and O1 as shown in Fig.~\ref{fig:dispersion} is fixed
manually to distinguish them.
}
\end{figure*}

\subsection{Scattering channel}         
\label{scatter}

The phonon scattering channels ruled by the conservation of energy are
investigated to quantify the specific scattering processes due to different
phonon branches.
Since the phonon branches commonly degenerate or cross with each other in the
most segments of the path passing through the main high-symmetry $k$-points in
the IBZ as shown in Fig.~\ref{fig:dispersion}, we only present in
Fig.~\ref{fig:scatter} the scattering rate of acoustic phonon modes along
$\Gamma$-Y direction where the acoustic phonon modes can be easily separated
as LA, TA and FA, and the crossover problem is fixed manually.
Both the scattering rates for absorption and emission processes as shown in
Eq.~(\ref{tau}) are addressed and the scattering rates for emission processes
are multiplied by 1/2 to avoid counting twice for the same process.
To focus on the scattering between acoustic and optical phonon modes, the LA, TA
and FA phonon branches are collected together as "A", the 3 low frequency
optical phonon branches are collected together as "O1", and the 6 optical phonon
branches with high frequency are collected together as "O2".
As discussed above, there exists a gap between O1 and O2 for monolayer $GeS$ and
$GeSe$, while the gap disappears for monolayer $SnS$ and $SnSe$.

All the possible phonon scattering channels for the four monolayer compounds
are plotted in Fig.~\ref{fig:scatter}.
Note that the phonon scattering channels are similar for $GeSe$, $SnS$ and
$SnSe$, while that for $GeS$ is different due to the lack of the scattering
channels of LA/TA/FA+A/O1$\to$O2, which is resulted from the large gap
($2.014\,\mathrm{THz}$) between O1 and O2 for monolayer $GeS$.
The major scattering channels that have a large contribution to the overall
scattering for monolayer $GeS$ are LA/TA/FA+A$\to$A, LA/TA/FA$\to$A+A and
LA+A$\to$O1.
In contrast, the major scattering channels for monolayer $GeSe$, $SnS$ and
$SnSe$ are LA/TA/FA+A$\to$O1.
The scattering between A and O1 phonon branches, especially for TA and FA, is
highly enhanced due to the tiny or disappeared gap for monolayer $GeSe$, $SnS$
and $SnSe$.
Furthermore, besides LA/TA/FA+A$\to$O1, the scattering channels of
TA/FA+O1$\to$O1 for monolayer $SnS$, and the scattering channels of
LA/TA/FA+O1$\to$O2, LA+A$\to$O2 and TA/FA+O2$\to$O2 for monolayer $SnSe$ play a
major role.
Along with the characteristics of their phonon dispersions as shown in
Fig.~\ref{fig:dispersion}, it can be concluded that a large gap will result in
the decoupling between acoustic and high frequency optical phonon modes, and
when the gap is tiny or disappears, the phonon branches becomes close to each
other, which will result in their strong couplings and lead to a remarkable
scattering channel.

It is discussed above based on Fig.~\ref{fig:kv} that the phonon transport
properties of the four monolayer compounds are nearly isotropic for high
frequency phonon modes above the separatrix frequency (where the gap
locates), which separates the phonon dispersion into two regions containing
equal numbers of branches, while the anisotropy along the two different
directions is conspicuous for low frequency phonon modes below the separatrix
frequency.
It is the phonon modes in the region below the separatrix frequency that result
in the anisotropy.
The coupling between the phonon branches will suppress the anisotropy of the low
frequency phonon modes when there is no gap.
When a gap exists, the coupling will be weak, leading to the presence of the
anisotropy.
If the gap is very large, the coupling disappears, resulting in the appearance
of the strong anisotropy of low frequency phonon modes.
Thus the anisotropy is huge for monolayer $GeS$, small for monolayer $GeSe$,
while almost disappear for monolayer $SnS$ and $SnSe$.

\subsection{Electron localization functions} 
\label{elf}

\begin{figure}[!tb]
\centering
    \includegraphics[width=0.46\textwidth]{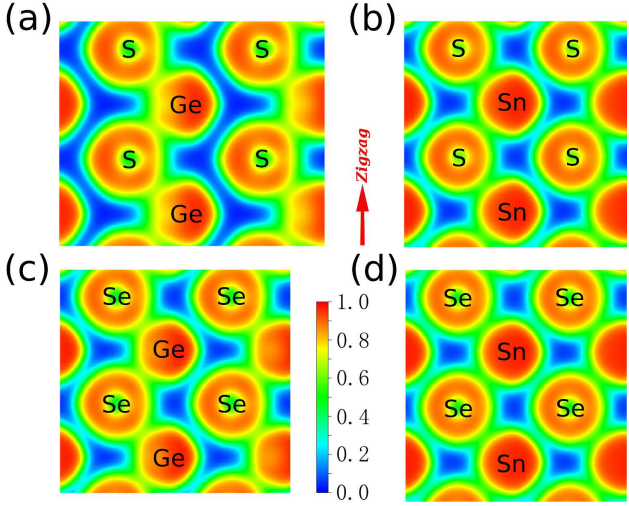}
\caption{\label{fig:ELF}
(Color online)
The top view of electron localization functions of monolayer (a) $GeS$, (b)
$SnS$, (c) $GeSe$ and (d) $SnSe$.  The symbols of atoms are marked on site.
}
\end{figure}

The anisotropic behavior can also be understood from a fundamental view of the
atomic bonding.
To picture the electron pair probability, we plot the ELF for the four monolayer
compounds in Fig.~\ref{fig:ELF}.
The ELF contains information on the structure of atomic shells, and also
displays the location and size of bonding and lone electron pairs.
\cite{Angew.Chem.Int.Ed.Engl..1997.36.17.1808-1832}
The ELF characterizes the probability of finding an electron with the same spin
in the neighborhood space of the reference electron.
The smaller the probability the more localized the reference electron.
The ELF is powerful in interpreting chemical bonding patterns, ranging from 0 to
1,\cite{C1CP21055F} where 0 means no electron, 0.5 corresponds to
electron-gas-like pair probability, and 1 corresponds to perfect localization.

As revealed in Fig.~\ref{fig:ELF}, the ELF along the zigzag direction for the
four monolayer compounds are all larger than 0.5 (the value of uniform electron
gas), which means that the electrons are localized.
The bonding makes the rigidity along zigzag direction, while for armchair
direction, the ELF are all smaller than 0.5, which means that the electrons are
delocalized.
The ELF along armchair direction is obviously the smallest for monolayer $GeS$,
resulting in its softest nature along the armchair direction reflected by its
smallest Young's modulus as shown in Table.~\ref{tab:collect}.
The ELF along the two directions shows a big difference for monolayer $GeS$ and
$GeSe$, especially for monolayer $GeS$, while shows only a small difference for
monolayer $SnS$ and $SnSe$.
The anisotropic behavior of the ELF reflecting the bonding characteristics is
the physical origin of the diverse anisotropic properties for the series of
monolayer compounds with hinge-like orthorhombic structure.

\section{Size effect}                   
\label{size}

It is well known that, with a larger rMFP, the thermal conductivity could be
modulated more effectively by nanostructuring.
If the rMFP is very small, the effect of nanostructures with a typical size will
be not significant for the phonon transport.
As listed in Table~\ref{tab:collect}, bulk $SnSe$ has the smallest rMFP, and the
four monolayer compounds have larger rMFP than bulk $SnSe$, which means that it
would be more effective to modulate the phonon transport in the monolayer
compounds by nanostructuring.
Furthermore, the phonon transport in monolayer $GeS$ and $GeSe$ could be more
effectively modulated due to their larger rMFP compared to that in monolayer
$SnS$ and $SnSe$.
In order to address the size effect on the phonon transport properties, the
phonon boundary scattering due to the finite size should be considered, which
can be estimated by the standard equation:
\cite{ziman1961electrons, srivastava1990physics, PhysRevB.79.155413}
\begin{equation}
\frac{1}{\tau^{B}_{\lambda}} = \frac{1-p}{1+p}\frac{|v_{\lambda}|}{L},
\end{equation}
where $p$ is the specularity parameter, which means the fraction of specularly
scattered phonons depending on the roughness of the edge, ranging from 0 for a
completely rough edge to 1 for a perfectly smooth edge, $v_{\lambda}$ is the
phonon group velocity of the phonon mode $\lambda$, and $L$ is the system size
usually ranging from nanometers to micrometers.
The temperature gradient is assumed to be along the direction of the finite
sample length.
Anisotropy due to finite sample size is ignored because of the relatively large
sizes considered.
\cite{PhysRevB.89.155426, PhysRevB.83.235428}
Then the scattering rate of each phonon mode is calculated by the Matthiessen
rule:
\begin{equation}
\frac{1}{\tau_{\lambda}} = \frac{1}{\tau^{anh}_{\lambda}}
            + \frac{1}{\tau^{iso}_{\lambda}}
            + \frac{1}{\tau^{B}_{\lambda}},
\end{equation}
where $1/\tau^{anh}_{\lambda}$ is the intrinsic anharmonic scattering rates due
to phonon-phonon interactions, and $1/\tau^{iso}_{\lambda}$ is the scattering
rate due to isotopic impurity.

\begin{figure}[!tb]
\centering
    \includegraphics[width=0.46\textwidth]{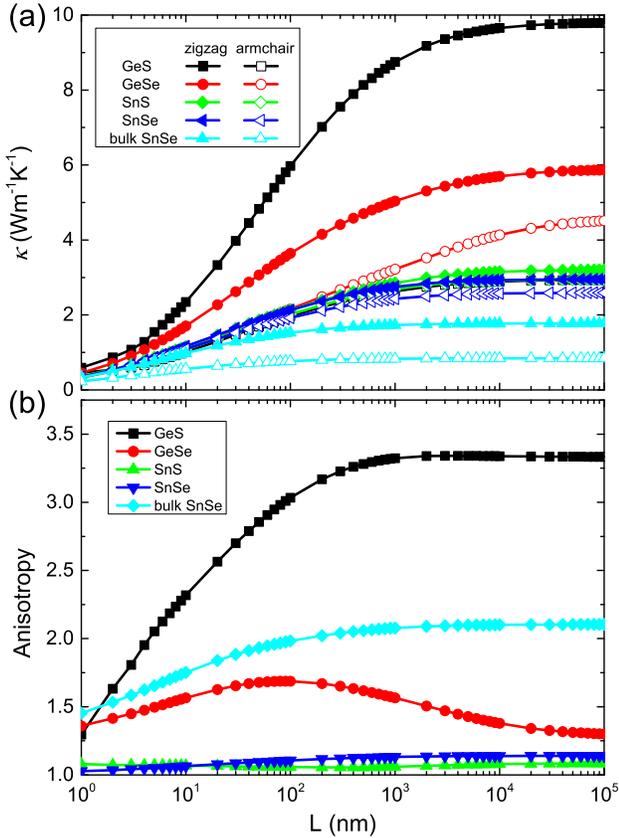}
\caption{\label{fig:boundary}
(Color online)
(a) The lattice thermal conductivity ($\kappa$) and (b) its anisotropy at
$300\,\mathrm{K}$ as a function of the system size for completely rough edges.
The anisotropy is defined as
$\kappa_{\mathrm{zigzag}}/\kappa_{\mathrm{armchair}}$.
Points are from calculations and lines are just guide to the eye.
}
\end{figure}

By considering a fully diffusive scattering at the boundary, i.e., $p=0$, we
plot the $\kappa$ as a function of the system size as shown in
Fig.~\ref{fig:boundary}(a).
When the size is up to near $10^5\,\mathrm{nm}$, the $\kappa$ for all the
monolayer and bulk compounds almost converge to the value of infinite system.
The $\kappa$ decreases with the size decreasing, and the $\kappa$ along armchair
direction of monolayer $GeSe$ has the most rapid decrease, which is resulted
from its largest rMFP of $72.51\,\mathrm{nm}$.
It is obvious that when the size decreases the $\kappa$ will decrease faster
with a larger rMFP.
It is also shown that the anisotropy of the $\kappa$ is affected by the finite
size.
Generally, the anisotropy will become weak due to the limited size.
The reason lies in that usually large $\kappa$ corresponds to large rMFP and
small $\kappa$ corresponds to small rMFP.
Thus the large $\kappa$ will have a bigger decrease than small $\kappa$ with
limited size, leading to the anisotropy of the $\kappa$ becoming weak at short
length.
Based on the $\kappa$ and rMFP as listed in Table.~\ref{tab:collect}, the
anisotropy of the $\kappa$ will become weak for monolayer $GeS$, $SnSe$ and bulk
$SnSe$ with limited size, while the contrary is the case for monolayer $GeSe$
and $SnS$ that the anisotropy of the $\kappa$ will become strong with limited
size.
The cases are confirmed as shown in Fig.~\ref{fig:boundary}(b).
With limited size, the $\kappa$ could be effectively lowered and the anisotropy
could be effectively modulated by nanostructuring such as patterning into
nanoribbon or incorporating pores, which would extend the applications in
thermoelectrics and thermal management.
Note that the size effect studied here is estimated based on diffusive phonon
transport by considering additional phonon boundary scattering.
If the ballistic phonon transport is effectively involved, the thermal
conductivity and its anisotropy might be slightly different from current results
at very small scales.
\cite{PhysicaE.2015.66..1-6}
For detailed study on the phonon transport properties of nanostructured systems,
more effective simulations should be performed, such as molecular dynamics (MD)
simulation and kinetic Monte Carlo (KMC).

\section{Summary and conclusions}       
\label{summary}

In summary, we have systematically investigated the diverse phonon transport
properties of 2D orthorhombic group IV-VI compounds of $GeS$, $GeSe$, $SnS$ and
$SnSe$ by solving the Boltzmann transport equation based on first-principles
calculations.
They all possess rather low thermal conductivity compared to lots of other 2D
materials.
The contribution from each phonon branch is studied and its relation with the
rMFP is also discussed.
The four monolayer compounds, although possessing similar hinge-like structure
along the armchair direction as phosphorene, show diverse anisotropic properties
in many aspects, such as phonon group velocity, Young's modulus and lattice
thermal conductivity ($\kappa$), etc.
Especially, the $\kappa$ along the zigzag and armchair directions of monolayer
$GeS$ shows the strongest anisotropy while monolayer $SnS$ and $SnSe$ shows an
almost isotropy in phonon transport, despite the similar characteristics of
their structures.
A detailed analysis on the diverse anisotropic properties of the series 2D
orthorhombic compounds is presented.
The anisotropy of thermal conductivity is actually dominated by the anisotropy
of phonon group velocity.
Based on the analysis of the frequency dependent $\kappa$ and average phonon
group velocity, we find that the anisotropy is mainly introduced by the region
below the gap between the high frequency optical phonon modes and low frequency
phonon modes in the phonon dispersions.
The larger the gap, the stronger the anisotropy.
The influence of the gap on the anisotropy can be explained from the coupling
between the phonon modes at both sides of the gap, which is supported by the
phonon scattering channels.
The diverse anisotropic behavior can also be understood from a fundamental view
of the atomic bonding characterized by the ELF.
The anisotropic behavior of the ELF reflecting the bonding characteristics is
the physical origin of the anisotropic properties.
The phonon transport in monolayer $GeS$ and $GeSe$ can be more effectively
modulated by nanostructuring due to their larger rMFP compared to that in
monolayer $SnS$ and $SnSe$.
It is also shown that the anisotropy of the $\kappa$ is affected by the finite
size.
With limited size, the $\kappa$ could be effectively lowered, and the anisotropy
could be effectively modulated by nanostructuring, which would extend the
applications in nanoscale thermoelectrics and thermal management.
This study not only present comprehensive investigations on the phonon transport
properties of the new family of 2D orthorhombic group IV-VI compounds ($GeS$,
$GeSe$, $SnS$ and $SnSe$), but also provide discussions and analysis on the
origins of the diverse anisotropy, which enriches the understanding of nanoscale
phonon transport in 2D materials, and would be of significance for further study
and applications in emerging technologies.

\section{Acknowledgments}               

The authors would like to thank Prof.\ Zhen-Gang Zhu and Prof.\ Qing-Rong Zheng
(University of Chinese Academy of Sciences, China) for helpful discussions.
G.Q.\ would like to thank Dr.\ J.\ Carrete and Prof.\ N.\ Mingo (CEA-Grenoble,
France) for providing data of bulk $SnSe$.
G.Q.\ and M.H.\ acknowledge the support by the Deutsche Forschungsgemeinschaft
(DFG) (project number: HU 2269/2-1).
This work is supported in part by the NSFC (Grant No.\ 11004239), the MOST
(Grant No.\ 2012CB932901) of China, and the fund from CAS.
The calculations were performed on Nebulae (DAWN6000) in National Supercomputing
Center in Shenzhen and MagicCube (DAWN5000A) in Shanghai Supercomputer Center,
China, and with computing resources granted by the J\"{u}lich Aachen Research
Alliance-High Performance Computing (JARA-HPC) from RWTH Aachen University under
Project No.\ jara0132.


\begin{thebibliography}{67}%
\makeatletter
\providecommand \@ifxundefined [1]{%
 \@ifx{#1\undefined}
}%
\providecommand \@ifnum [1]{%
 \ifnum #1\expandafter \@firstoftwo
 \else \expandafter \@secondoftwo
 \fi
}%
\providecommand \@ifx [1]{%
 \ifx #1\expandafter \@firstoftwo
 \else \expandafter \@secondoftwo
 \fi
}%
\providecommand \natexlab [1]{#1}%
\providecommand \enquote  [1]{``#1''}%
\providecommand \bibnamefont  [1]{#1}%
\providecommand \bibfnamefont [1]{#1}%
\providecommand \citenamefont [1]{#1}%
\providecommand \href@noop [0]{\@secondoftwo}%
\providecommand \href [0]{\begingroup \@sanitize@url \@href}%
\providecommand \@href[1]{\@@startlink{#1}\@@href}%
\providecommand \@@href[1]{\endgroup#1\@@endlink}%
\providecommand \@sanitize@url [0]{\catcode `\\12\catcode `\$12\catcode
  `\&12\catcode `\#12\catcode `\^12\catcode `\_12\catcode `\%12\relax}%
\providecommand \@@startlink[1]{}%
\providecommand \@@endlink[0]{}%
\providecommand \url  [0]{\begingroup\@sanitize@url \@url }%
\providecommand \@url [1]{\endgroup\@href {#1}{\urlprefix }}%
\providecommand \urlprefix  [0]{URL }%
\providecommand \Eprint [0]{\href }%
\providecommand \doibase [0]{http://dx.doi.org/}%
\providecommand \selectlanguage [0]{\@gobble}%
\providecommand \bibinfo  [0]{\@secondoftwo}%
\providecommand \bibfield  [0]{\@secondoftwo}%
\providecommand \translation [1]{[#1]}%
\providecommand \BibitemOpen [0]{}%
\providecommand \bibitemStop [0]{}%
\providecommand \bibitemNoStop [0]{.\EOS\space}%
\providecommand \EOS [0]{\spacefactor3000\relax}%
\providecommand \BibitemShut  [1]{\csname bibitem#1\endcsname}%
\let\auto@bib@innerbib\@empty
\bibitem [{\citenamefont {Zhao}\ \emph {et~al.}(2014)\citenamefont {Zhao},
  \citenamefont {Lo}, \citenamefont {Zhang}, \citenamefont {Sun}, \citenamefont
  {Tan}, \citenamefont {Uher}, \citenamefont {Wolverton}, \citenamefont
  {Dravid},\ and\ \citenamefont {Kanatzidis}}]{Nature.2014.508.7496.373-377}%
  \BibitemOpen
  \bibfield  {author} {\bibinfo {author} {\bibfnamefont {L.-D.}\ \bibnamefont
  {Zhao}}, \bibinfo {author} {\bibfnamefont {S.-H.}\ \bibnamefont {Lo}},
  \bibinfo {author} {\bibfnamefont {Y.}~\bibnamefont {Zhang}}, \bibinfo
  {author} {\bibfnamefont {H.}~\bibnamefont {Sun}}, \bibinfo {author}
  {\bibfnamefont {G.}~\bibnamefont {Tan}}, \bibinfo {author} {\bibfnamefont
  {C.}~\bibnamefont {Uher}}, \bibinfo {author} {\bibfnamefont {C.}~\bibnamefont
  {Wolverton}}, \bibinfo {author} {\bibfnamefont {V.~P.}\ \bibnamefont
  {Dravid}}, \ and\ \bibinfo {author} {\bibfnamefont {M.~G.}\ \bibnamefont
  {Kanatzidis}},\ }\href {\doibase 10.1038/nature13184} {\bibfield  {journal}
  {\bibinfo  {journal} {Nature}\ }\textbf {\bibinfo {volume} {508}},\ \bibinfo
  {pages} {373} (\bibinfo {year} {2014})}\BibitemShut {NoStop}%
\bibitem [{\citenamefont {Zhao}\ \emph {et~al.}(2016)\citenamefont {Zhao},
  \citenamefont {Tan}, \citenamefont {Hao}, \citenamefont {He}, \citenamefont
  {Pei}, \citenamefont {Chi}, \citenamefont {Wang}, \citenamefont {Gong},
  \citenamefont {Xu},\ and\ \citenamefont
  {Dravid}}]{Science.2016.351.6269.141-144}%
  \BibitemOpen
  \bibfield  {author} {\bibinfo {author} {\bibfnamefont {L.-D.}\ \bibnamefont
  {Zhao}}, \bibinfo {author} {\bibfnamefont {G.}~\bibnamefont {Tan}}, \bibinfo
  {author} {\bibfnamefont {S.}~\bibnamefont {Hao}}, \bibinfo {author}
  {\bibfnamefont {J.}~\bibnamefont {He}}, \bibinfo {author} {\bibfnamefont
  {Y.}~\bibnamefont {Pei}}, \bibinfo {author} {\bibfnamefont {H.}~\bibnamefont
  {Chi}}, \bibinfo {author} {\bibfnamefont {H.}~\bibnamefont {Wang}}, \bibinfo
  {author} {\bibfnamefont {S.}~\bibnamefont {Gong}}, \bibinfo {author}
  {\bibfnamefont {H.}~\bibnamefont {Xu}}, \ and\ \bibinfo {author}
  {\bibfnamefont {V.~P.}\ \bibnamefont {Dravid}},\ }\href@noop {} {\bibfield
  {journal} {\bibinfo  {journal} {Science}\ }\textbf {\bibinfo {volume}
  {351}},\ \bibinfo {pages} {141} (\bibinfo {year} {2016})}\BibitemShut
  {NoStop}%
\bibitem [{\citenamefont {Parenteau}\ and\ \citenamefont
  {Carlone}(1990)}]{PhysRevB.41.5227}%
  \BibitemOpen
  \bibfield  {author} {\bibinfo {author} {\bibfnamefont {M.}~\bibnamefont
  {Parenteau}}\ and\ \bibinfo {author} {\bibfnamefont {C.}~\bibnamefont
  {Carlone}},\ }\href {\doibase 10.1103/PhysRevB.41.5227} {\bibfield  {journal}
  {\bibinfo  {journal} {Phys. Rev. B}\ }\textbf {\bibinfo {volume} {41}},\
  \bibinfo {pages} {5227} (\bibinfo {year} {1990})}\BibitemShut {NoStop}%
\bibitem [{\citenamefont {Guo}\ \emph {et~al.}(2015{\natexlab{a}})\citenamefont
  {Guo}, \citenamefont {Wang}, \citenamefont {Kuang},\ and\ \citenamefont
  {Huang}}]{PhysRevB.92.115202}%
  \BibitemOpen
  \bibfield  {author} {\bibinfo {author} {\bibfnamefont {R.}~\bibnamefont
  {Guo}}, \bibinfo {author} {\bibfnamefont {X.}~\bibnamefont {Wang}}, \bibinfo
  {author} {\bibfnamefont {Y.}~\bibnamefont {Kuang}}, \ and\ \bibinfo {author}
  {\bibfnamefont {B.}~\bibnamefont {Huang}},\ }\href {\doibase
  10.1103/PhysRevB.92.115202} {\bibfield  {journal} {\bibinfo  {journal} {Phys.
  Rev. B}\ }\textbf {\bibinfo {volume} {92}},\ \bibinfo {pages} {115202}
  (\bibinfo {year} {2015}{\natexlab{a}})}\BibitemShut {NoStop}%
\bibitem [{\citenamefont {Carrete}\ \emph {et~al.}(2014)\citenamefont
  {Carrete}, \citenamefont {Mingo},\ and\ \citenamefont
  {Curtarolo}}]{apl.10510101907.1.4895770}%
  \BibitemOpen
  \bibfield  {author} {\bibinfo {author} {\bibfnamefont {J.}~\bibnamefont
  {Carrete}}, \bibinfo {author} {\bibfnamefont {N.}~\bibnamefont {Mingo}}, \
  and\ \bibinfo {author} {\bibfnamefont {S.}~\bibnamefont {Curtarolo}},\ }\href
  {\doibase http://dx.doi.org/10.1063/1.4895770} {\bibfield  {journal}
  {\bibinfo  {journal} {Appl. Phys. Lett.}\ }\textbf {\bibinfo {volume}
  {105}},\ \bibinfo {eid} {101907} (\bibinfo {year} {2014})}\BibitemShut
  {NoStop}%
\bibitem [{\citenamefont {Shi}\ and\ \citenamefont
  {Kioupakis}(2015)}]{jap.1.4907805}%
  \BibitemOpen
  \bibfield  {author} {\bibinfo {author} {\bibfnamefont {G.}~\bibnamefont
  {Shi}}\ and\ \bibinfo {author} {\bibfnamefont {E.}~\bibnamefont
  {Kioupakis}},\ }\href {\doibase http://dx.doi.org/10.1063/1.4907805}
  {\bibfield  {journal} {\bibinfo  {journal} {J. Appl. Phys.}\ }\textbf
  {\bibinfo {volume} {117}},\ \bibinfo {eid} {065103} (\bibinfo {year}
  {2015})}\BibitemShut {NoStop}%
\bibitem [{\citenamefont {Chen}\ \emph {et~al.}(2014)\citenamefont {Chen},
  \citenamefont {Wang}, \citenamefont {Chen}, \citenamefont {Day},\ and\
  \citenamefont {Snyder}}]{C4TA01643B}%
  \BibitemOpen
  \bibfield  {author} {\bibinfo {author} {\bibfnamefont {C.-L.}\ \bibnamefont
  {Chen}}, \bibinfo {author} {\bibfnamefont {H.}~\bibnamefont {Wang}}, \bibinfo
  {author} {\bibfnamefont {Y.-Y.}\ \bibnamefont {Chen}}, \bibinfo {author}
  {\bibfnamefont {T.}~\bibnamefont {Day}}, \ and\ \bibinfo {author}
  {\bibfnamefont {G.~J.}\ \bibnamefont {Snyder}},\ }\href {\doibase
  10.1039/C4TA01643B} {\bibfield  {journal} {\bibinfo  {journal} {J. Mater.
  Chem. A}\ }\textbf {\bibinfo {volume} {2}},\ \bibinfo {pages} {11171}
  (\bibinfo {year} {2014})}\BibitemShut {NoStop}%
\bibitem [{\citenamefont {Tan}\ \emph {et~al.}(2014)\citenamefont {Tan},
  \citenamefont {Zhao}, \citenamefont {Li}, \citenamefont {Wu}, \citenamefont
  {Wei}, \citenamefont {Xing},\ and\ \citenamefont {Kanatzidis}}]{C4TA04462B}%
  \BibitemOpen
  \bibfield  {author} {\bibinfo {author} {\bibfnamefont {Q.}~\bibnamefont
  {Tan}}, \bibinfo {author} {\bibfnamefont {L.-D.}\ \bibnamefont {Zhao}},
  \bibinfo {author} {\bibfnamefont {J.-F.}\ \bibnamefont {Li}}, \bibinfo
  {author} {\bibfnamefont {C.-F.}\ \bibnamefont {Wu}}, \bibinfo {author}
  {\bibfnamefont {T.-R.}\ \bibnamefont {Wei}}, \bibinfo {author} {\bibfnamefont
  {Z.-B.}\ \bibnamefont {Xing}}, \ and\ \bibinfo {author} {\bibfnamefont
  {M.~G.}\ \bibnamefont {Kanatzidis}},\ }\href {\doibase 10.1039/C4TA04462B}
  {\bibfield  {journal} {\bibinfo  {journal} {J. Mater. Chem. A}\ }\textbf
  {\bibinfo {volume} {2}},\ \bibinfo {pages} {17302} (\bibinfo {year}
  {2014})}\BibitemShut {NoStop}%
\bibitem [{\citenamefont {Zhu}\ \emph {et~al.}(2014)\citenamefont {Zhu},
  \citenamefont {Sun}, \citenamefont {Armiento}, \citenamefont {Lazic},\ and\
  \citenamefont {Ceder}}]{apl.1.4866861}%
  \BibitemOpen
  \bibfield  {author} {\bibinfo {author} {\bibfnamefont {H.}~\bibnamefont
  {Zhu}}, \bibinfo {author} {\bibfnamefont {W.}~\bibnamefont {Sun}}, \bibinfo
  {author} {\bibfnamefont {R.}~\bibnamefont {Armiento}}, \bibinfo {author}
  {\bibfnamefont {P.}~\bibnamefont {Lazic}}, \ and\ \bibinfo {author}
  {\bibfnamefont {G.}~\bibnamefont {Ceder}},\ }\href {\doibase
  http://dx.doi.org/10.1063/1.4866861} {\bibfield  {journal} {\bibinfo
  {journal} {Appl. Phys. Lett.}\ }\textbf {\bibinfo {volume} {104}},\ \bibinfo
  {eid} {082107} (\bibinfo {year} {2014})}\BibitemShut {NoStop}%
\bibitem [{\citenamefont {Chandrasekhar}\ \emph {et~al.}(1977)\citenamefont
  {Chandrasekhar}, \citenamefont {Humphreys}, \citenamefont {Zwick},\ and\
  \citenamefont {Cardona}}]{PhysRevB.15.2177}%
  \BibitemOpen
  \bibfield  {author} {\bibinfo {author} {\bibfnamefont {H.~R.}\ \bibnamefont
  {Chandrasekhar}}, \bibinfo {author} {\bibfnamefont {R.~G.}\ \bibnamefont
  {Humphreys}}, \bibinfo {author} {\bibfnamefont {U.}~\bibnamefont {Zwick}}, \
  and\ \bibinfo {author} {\bibfnamefont {M.}~\bibnamefont {Cardona}},\ }\href
  {\doibase 10.1103/PhysRevB.15.2177} {\bibfield  {journal} {\bibinfo
  {journal} {Phys. Rev. B}\ }\textbf {\bibinfo {volume} {15}},\ \bibinfo
  {pages} {2177} (\bibinfo {year} {1977})}\BibitemShut {NoStop}%
\bibitem [{\citenamefont {Reijnders}\ \emph {et~al.}(2014)\citenamefont
  {Reijnders}, \citenamefont {Hamilton}, \citenamefont {Britto}, \citenamefont
  {Brubach}, \citenamefont {Roy}, \citenamefont {Gibson}, \citenamefont
  {Cava},\ and\ \citenamefont {Burch}}]{PhysRevB.90.235144}%
  \BibitemOpen
  \bibfield  {author} {\bibinfo {author} {\bibfnamefont {A.~A.}\ \bibnamefont
  {Reijnders}}, \bibinfo {author} {\bibfnamefont {J.}~\bibnamefont {Hamilton}},
  \bibinfo {author} {\bibfnamefont {V.}~\bibnamefont {Britto}}, \bibinfo
  {author} {\bibfnamefont {J.-B.}\ \bibnamefont {Brubach}}, \bibinfo {author}
  {\bibfnamefont {P.}~\bibnamefont {Roy}}, \bibinfo {author} {\bibfnamefont
  {Q.~D.}\ \bibnamefont {Gibson}}, \bibinfo {author} {\bibfnamefont {R.~J.}\
  \bibnamefont {Cava}}, \ and\ \bibinfo {author} {\bibfnamefont {K.~S.}\
  \bibnamefont {Burch}},\ }\href {\doibase 10.1103/PhysRevB.90.235144}
  {\bibfield  {journal} {\bibinfo  {journal} {Phys. Rev. B}\ }\textbf {\bibinfo
  {volume} {90}},\ \bibinfo {pages} {235144} (\bibinfo {year}
  {2014})}\BibitemShut {NoStop}%
\bibitem [{\citenamefont {Rao}\ and\ \citenamefont
  {Chaudhuri}(1985)}]{0022-3727-18-6-003}%
  \BibitemOpen
  \bibfield  {author} {\bibinfo {author} {\bibfnamefont {T.~S.}\ \bibnamefont
  {Rao}}\ and\ \bibinfo {author} {\bibfnamefont {A.~K.}\ \bibnamefont
  {Chaudhuri}},\ }\href {http://stacks.iop.org/0022-3727/18/i=6/a=003}
  {\bibfield  {journal} {\bibinfo  {journal} {J. Phys. D: Appl. Phys.}\
  }\textbf {\bibinfo {volume} {18}},\ \bibinfo {pages} {L35} (\bibinfo {year}
  {1985})}\BibitemShut {NoStop}%
\bibitem [{\citenamefont {Baumgardner}\ \emph {et~al.}(2010)\citenamefont
  {Baumgardner}, \citenamefont {Choi}, \citenamefont {Lim},\ and\ \citenamefont
  {Hanrath}}]{J.Am.Chem.Soc..2010.132.28.9519-9521}%
  \BibitemOpen
  \bibfield  {author} {\bibinfo {author} {\bibfnamefont {W.~J.}\ \bibnamefont
  {Baumgardner}}, \bibinfo {author} {\bibfnamefont {J.~J.}\ \bibnamefont
  {Choi}}, \bibinfo {author} {\bibfnamefont {Y.-F.}\ \bibnamefont {Lim}}, \
  and\ \bibinfo {author} {\bibfnamefont {T.}~\bibnamefont {Hanrath}},\ }\href
  {\doibase 10.1021/ja1013745} {\bibfield  {journal} {\bibinfo  {journal} {J.
  Am. Chem. Soc.}\ }\textbf {\bibinfo {volume} {132}},\ \bibinfo {pages} {9519}
  (\bibinfo {year} {2010})}\BibitemShut {NoStop}%
\bibitem [{\citenamefont {Gomes}\ and\ \citenamefont
  {Carvalho}(2015)}]{PhysRevB.92.085406}%
  \BibitemOpen
  \bibfield  {author} {\bibinfo {author} {\bibfnamefont {L.~C.}\ \bibnamefont
  {Gomes}}\ and\ \bibinfo {author} {\bibfnamefont {A.}~\bibnamefont
  {Carvalho}},\ }\href {\doibase 10.1103/PhysRevB.92.085406} {\bibfield
  {journal} {\bibinfo  {journal} {Phys. Rev. B}\ }\textbf {\bibinfo {volume}
  {92}},\ \bibinfo {pages} {085406} (\bibinfo {year} {2015})}\BibitemShut
  {NoStop}%
\bibitem [{\citenamefont {Vidal}\ \emph {et~al.}(2012)\citenamefont {Vidal},
  \citenamefont {Lany}, \citenamefont {d'Avezac}, \citenamefont {Zunger},
  \citenamefont {Zakutayev}, \citenamefont {Francis},\ and\ \citenamefont
  {Tate}}]{AppliedPhysicsLetters.2012.100.3.032104}%
  \BibitemOpen
  \bibfield  {author} {\bibinfo {author} {\bibfnamefont {J.}~\bibnamefont
  {Vidal}}, \bibinfo {author} {\bibfnamefont {S.}~\bibnamefont {Lany}},
  \bibinfo {author} {\bibfnamefont {M.}~\bibnamefont {d'Avezac}}, \bibinfo
  {author} {\bibfnamefont {A.}~\bibnamefont {Zunger}}, \bibinfo {author}
  {\bibfnamefont {A.}~\bibnamefont {Zakutayev}}, \bibinfo {author}
  {\bibfnamefont {J.}~\bibnamefont {Francis}}, \ and\ \bibinfo {author}
  {\bibfnamefont {J.}~\bibnamefont {Tate}},\ }\href {\doibase
  http://dx.doi.org/10.1063/1.3675880} {\bibfield  {journal} {\bibinfo
  {journal} {Appl. Phys. Lett.}\ }\textbf {\bibinfo {volume} {100}},\ \bibinfo
  {pages} {032104} (\bibinfo {year} {2012})}\BibitemShut {NoStop}%
\bibitem [{\citenamefont {Cahill}\ \emph {et~al.}(1992)\citenamefont {Cahill},
  \citenamefont {Watson},\ and\ \citenamefont {Pohl}}]{PhysRevB.46.6131}%
  \BibitemOpen
  \bibfield  {author} {\bibinfo {author} {\bibfnamefont {D.~G.}\ \bibnamefont
  {Cahill}}, \bibinfo {author} {\bibfnamefont {S.~K.}\ \bibnamefont {Watson}},
  \ and\ \bibinfo {author} {\bibfnamefont {R.~O.}\ \bibnamefont {Pohl}},\
  }\href {\doibase 10.1103/PhysRevB.46.6131} {\bibfield  {journal} {\bibinfo
  {journal} {Phys. Rev. B}\ }\textbf {\bibinfo {volume} {46}},\ \bibinfo
  {pages} {6131} (\bibinfo {year} {1992})}\BibitemShut {NoStop}%
\bibitem [{\citenamefont {Ding}\ \emph
  {et~al.}(2015{\natexlab{a}})\citenamefont {Ding}, \citenamefont {Gao},\ and\
  \citenamefont {Yao}}]{ScientificReports.2015.5..9567}%
  \BibitemOpen
  \bibfield  {author} {\bibinfo {author} {\bibfnamefont {G.}~\bibnamefont
  {Ding}}, \bibinfo {author} {\bibfnamefont {G.}~\bibnamefont {Gao}}, \ and\
  \bibinfo {author} {\bibfnamefont {K.}~\bibnamefont {Yao}},\ }\href {\doibase
  10.1038/srep09567} {\bibfield  {journal} {\bibinfo  {journal} {Sci. Rep.}\
  }\textbf {\bibinfo {volume} {5}},\ \bibinfo {pages} {9567} (\bibinfo {year}
  {2015}{\natexlab{a}})}\BibitemShut {NoStop}%
\bibitem [{\citenamefont {Zhang}\ \emph {et~al.}(2015)\citenamefont {Zhang},
  \citenamefont {Bao},\ and\ \citenamefont {Hu}}]{C4NR06523A}%
  \BibitemOpen
  \bibfield  {author} {\bibinfo {author} {\bibfnamefont {X.}~\bibnamefont
  {Zhang}}, \bibinfo {author} {\bibfnamefont {H.}~\bibnamefont {Bao}}, \ and\
  \bibinfo {author} {\bibfnamefont {M.}~\bibnamefont {Hu}},\ }\href {\doibase
  10.1039/C4NR06523A} {\bibfield  {journal} {\bibinfo  {journal} {Nanoscale}\
  }\textbf {\bibinfo {volume} {7}},\ \bibinfo {pages} {6014} (\bibinfo {year}
  {2015})}\BibitemShut {NoStop}%
\bibitem [{\citenamefont {Gao}\ \emph {et~al.}(2015)\citenamefont {Gao},
  \citenamefont {Zhang}, \citenamefont {Jing},\ and\ \citenamefont
  {Hu}}]{Nanoscale.2015.7.16.7143-7150}%
  \BibitemOpen
  \bibfield  {author} {\bibinfo {author} {\bibfnamefont {Y.}~\bibnamefont
  {Gao}}, \bibinfo {author} {\bibfnamefont {X.}~\bibnamefont {Zhang}}, \bibinfo
  {author} {\bibfnamefont {Y.}~\bibnamefont {Jing}}, \ and\ \bibinfo {author}
  {\bibfnamefont {M.}~\bibnamefont {Hu}},\ }\href {\doibase 10.1039/C4NR07359B}
  {\bibfield  {journal} {\bibinfo  {journal} {Nanoscale}\ }\textbf {\bibinfo
  {volume} {7}},\ \bibinfo {pages} {7143} (\bibinfo {year} {2015})}\BibitemShut
  {NoStop}%
\bibitem [{\citenamefont {Xu}\ \emph {et~al.}(2016)\citenamefont {Xu},
  \citenamefont {Yang}, \citenamefont {Zhu}, \citenamefont {Yan}, \citenamefont
  {Pei}, \citenamefont {Myint}, \citenamefont {Zhang},\ and\ \citenamefont
  {Lu}}]{Nanoscale.2016.8.1.129-135}%
  \BibitemOpen
  \bibfield  {author} {\bibinfo {author} {\bibfnamefont {R.}~\bibnamefont
  {Xu}}, \bibinfo {author} {\bibfnamefont {J.}~\bibnamefont {Yang}}, \bibinfo
  {author} {\bibfnamefont {Y.}~\bibnamefont {Zhu}}, \bibinfo {author}
  {\bibfnamefont {H.}~\bibnamefont {Yan}}, \bibinfo {author} {\bibfnamefont
  {J.}~\bibnamefont {Pei}}, \bibinfo {author} {\bibfnamefont {Y.~W.}\
  \bibnamefont {Myint}}, \bibinfo {author} {\bibfnamefont {S.}~\bibnamefont
  {Zhang}}, \ and\ \bibinfo {author} {\bibfnamefont {Y.}~\bibnamefont {Lu}},\
  }\href {\doibase 10.1039/C5NR04366B} {\bibfield  {journal} {\bibinfo
  {journal} {Nanoscale}\ }\textbf {\bibinfo {volume} {8}},\ \bibinfo {pages}
  {129} (\bibinfo {year} {2016})}\BibitemShut {NoStop}%
\bibitem [{\citenamefont {Hong}\ \emph {et~al.}(2015)\citenamefont {Hong},
  \citenamefont {Zhang}, \citenamefont {Huang},\ and\ \citenamefont
  {Zeng}}]{Nanoscale.2015.7.44.18716-18724}%
  \BibitemOpen
  \bibfield  {author} {\bibinfo {author} {\bibfnamefont {Y.}~\bibnamefont
  {Hong}}, \bibinfo {author} {\bibfnamefont {J.}~\bibnamefont {Zhang}},
  \bibinfo {author} {\bibfnamefont {X.}~\bibnamefont {Huang}}, \ and\ \bibinfo
  {author} {\bibfnamefont {X.~C.}\ \bibnamefont {Zeng}},\ }\href {\doibase
  10.1039/C5NR03577E} {\bibfield  {journal} {\bibinfo  {journal} {Nanoscale}\
  }\textbf {\bibinfo {volume} {7}},\ \bibinfo {pages} {18716} (\bibinfo {year}
  {2015})}\BibitemShut {NoStop}%
\bibitem [{\citenamefont {Churchill}\ and\ \citenamefont
  {Jarillo-Herrero}(2014)}]{BP_NATURE}%
  \BibitemOpen
  \bibfield  {author} {\bibinfo {author} {\bibfnamefont {H.~O.~H.}\
  \bibnamefont {Churchill}}\ and\ \bibinfo {author} {\bibfnamefont
  {P.}~\bibnamefont {Jarillo-Herrero}},\ }\href
  {http://dx.doi.org/10.1038/nnano.2014.85} {\bibfield  {journal} {\bibinfo
  {journal} {Nature Nanotech.}\ }\textbf {\bibinfo {volume} {9}},\ \bibinfo
  {pages} {330} (\bibinfo {year} {2014})}\BibitemShut {NoStop}%
\bibitem [{\citenamefont {Li}\ \emph {et~al.}(2013{\natexlab{a}})\citenamefont
  {Li}, \citenamefont {Chen}, \citenamefont {Hu}, \citenamefont {Wang},
  \citenamefont {Zhang}, \citenamefont {Chen},\ and\ \citenamefont
  {Wang}}]{JACS.ja3108017}%
  \BibitemOpen
  \bibfield  {author} {\bibinfo {author} {\bibfnamefont {L.}~\bibnamefont
  {Li}}, \bibinfo {author} {\bibfnamefont {Z.}~\bibnamefont {Chen}}, \bibinfo
  {author} {\bibfnamefont {Y.}~\bibnamefont {Hu}}, \bibinfo {author}
  {\bibfnamefont {X.}~\bibnamefont {Wang}}, \bibinfo {author} {\bibfnamefont
  {T.}~\bibnamefont {Zhang}}, \bibinfo {author} {\bibfnamefont
  {W.}~\bibnamefont {Chen}}, \ and\ \bibinfo {author} {\bibfnamefont
  {Q.}~\bibnamefont {Wang}},\ }\href {\doibase 10.1021/ja3108017} {\bibfield
  {journal} {\bibinfo  {journal} {J. Am. Chem. Soc.}\ }\textbf {\bibinfo
  {volume} {135}},\ \bibinfo {pages} {1213} (\bibinfo {year}
  {2013}{\natexlab{a}})}\BibitemShut {NoStop}%
\bibitem [{\citenamefont {Antunez}\ \emph {et~al.}(2011)\citenamefont
  {Antunez}, \citenamefont {Buckley},\ and\ \citenamefont
  {Brutchey}}]{C1NR10084J}%
  \BibitemOpen
  \bibfield  {author} {\bibinfo {author} {\bibfnamefont {P.~D.}\ \bibnamefont
  {Antunez}}, \bibinfo {author} {\bibfnamefont {J.~J.}\ \bibnamefont
  {Buckley}}, \ and\ \bibinfo {author} {\bibfnamefont {R.~L.}\ \bibnamefont
  {Brutchey}},\ }\href {\doibase 10.1039/C1NR10084J} {\bibfield  {journal}
  {\bibinfo  {journal} {Nanoscale}\ }\textbf {\bibinfo {volume} {3}},\ \bibinfo
  {pages} {2399} (\bibinfo {year} {2011})}\BibitemShut {NoStop}%
\bibitem [{\citenamefont {Fei}\ \emph {et~al.}(2015)\citenamefont {Fei},
  \citenamefont {Li}, \citenamefont {Li},\ and\ \citenamefont
  {Yang}}]{AppliedPhysicsLetters.2015.107.17.173104}%
  \BibitemOpen
  \bibfield  {author} {\bibinfo {author} {\bibfnamefont {R.}~\bibnamefont
  {Fei}}, \bibinfo {author} {\bibfnamefont {W.}~\bibnamefont {Li}}, \bibinfo
  {author} {\bibfnamefont {J.}~\bibnamefont {Li}}, \ and\ \bibinfo {author}
  {\bibfnamefont {L.}~\bibnamefont {Yang}},\ }\href {\doibase
  http://dx.doi.org/10.1063/1.4934750} {\bibfield  {journal} {\bibinfo
  {journal} {Appl. Phys. Lett.}\ }\textbf {\bibinfo {volume} {107}},\ \bibinfo
  {pages} {173104} (\bibinfo {year} {2015})}\BibitemShut {NoStop}%
\bibitem [{\citenamefont {Wang}\ \emph {et~al.}(2015)\citenamefont {Wang},
  \citenamefont {Zhang}, \citenamefont {Yu},\ and\ \citenamefont
  {Wang}}]{Nanoscale.2015.7.38.15962-15970}%
  \BibitemOpen
  \bibfield  {author} {\bibinfo {author} {\bibfnamefont {F.~Q.}\ \bibnamefont
  {Wang}}, \bibinfo {author} {\bibfnamefont {S.}~\bibnamefont {Zhang}},
  \bibinfo {author} {\bibfnamefont {J.}~\bibnamefont {Yu}}, \ and\ \bibinfo
  {author} {\bibfnamefont {Q.}~\bibnamefont {Wang}},\ }\href {\doibase
  10.1039/C5NR03813H} {\bibfield  {journal} {\bibinfo  {journal} {Nanoscale}\
  }\textbf {\bibinfo {volume} {7}},\ \bibinfo {pages} {15962} (\bibinfo {year}
  {2015})}\BibitemShut {NoStop}%
\bibitem [{\citenamefont {Ding}\ \emph
  {et~al.}(2015{\natexlab{b}})\citenamefont {Ding}, \citenamefont {Gao} \emph
  {et~al.}}]{ding2015thermoelectric}%
  \BibitemOpen
  \bibfield  {author} {\bibinfo {author} {\bibfnamefont {G.}~\bibnamefont
  {Ding}}, \bibinfo {author} {\bibfnamefont {G.}~\bibnamefont {Gao}},  \emph
  {et~al.},\ }\href@noop {} {\bibfield  {journal} {\bibinfo  {journal} {arXiv
  preprint arXiv:1509.01759}\ } (\bibinfo {year}
  {2015}{\natexlab{b}})}\BibitemShut {NoStop}%
\bibitem [{\citenamefont {Zhang}\ \emph {et~al.}(2016)\citenamefont {Zhang},
  \citenamefont {Qin}, \citenamefont {Fang}, \citenamefont {Cui}, \citenamefont
  {Zheng}, \citenamefont {Yan},\ and\ \citenamefont
  {Su}}]{Sci.Rep..2016.6..19830}%
  \BibitemOpen
  \bibfield  {author} {\bibinfo {author} {\bibfnamefont {L.-C.}\ \bibnamefont
  {Zhang}}, \bibinfo {author} {\bibfnamefont {G.}~\bibnamefont {Qin}}, \bibinfo
  {author} {\bibfnamefont {W.-Z.}\ \bibnamefont {Fang}}, \bibinfo {author}
  {\bibfnamefont {H.-J.}\ \bibnamefont {Cui}}, \bibinfo {author} {\bibfnamefont
  {Q.-R.}\ \bibnamefont {Zheng}}, \bibinfo {author} {\bibfnamefont {Q.-B.}\
  \bibnamefont {Yan}}, \ and\ \bibinfo {author} {\bibfnamefont
  {G.}~\bibnamefont {Su}},\ }\href {\doibase 10.1038/srep19830} {\bibfield
  {journal} {\bibinfo  {journal} {Sci. Rep.}\ }\textbf {\bibinfo {volume}
  {6}},\ \bibinfo {pages} {19830} (\bibinfo {year} {2016})}\BibitemShut
  {NoStop}%
\bibitem [{\citenamefont {Safaei}\ \emph {et~al.}(2015)\citenamefont {Safaei},
  \citenamefont {Galicka}, \citenamefont {Kacman},\ and\ \citenamefont
  {Buczko}}]{safaei2015quantum}%
  \BibitemOpen
  \bibfield  {author} {\bibinfo {author} {\bibfnamefont {S.}~\bibnamefont
  {Safaei}}, \bibinfo {author} {\bibfnamefont {M.}~\bibnamefont {Galicka}},
  \bibinfo {author} {\bibfnamefont {P.}~\bibnamefont {Kacman}}, \ and\ \bibinfo
  {author} {\bibfnamefont {R.}~\bibnamefont {Buczko}},\ }\href@noop {}
  {\bibfield  {journal} {\bibinfo  {journal} {arXiv preprint arXiv:1508.01364}\
  } (\bibinfo {year} {2015})}\BibitemShut {NoStop}%
\bibitem [{\citenamefont {Fang}\ \emph {et~al.}(2016)\citenamefont {Fang},
  \citenamefont {Zhang}, \citenamefont {Qin}, \citenamefont {Yan},
  \citenamefont {Zheng},\ and\ \citenamefont {Su}}]{arXiv1603.01791.2016}%
  \BibitemOpen
  \bibfield  {author} {\bibinfo {author} {\bibfnamefont {W.-Z.}\ \bibnamefont
  {Fang}}, \bibinfo {author} {\bibfnamefont {L.-C.}\ \bibnamefont {Zhang}},
  \bibinfo {author} {\bibfnamefont {G.}~\bibnamefont {Qin}}, \bibinfo {author}
  {\bibfnamefont {Q.-B.}\ \bibnamefont {Yan}}, \bibinfo {author} {\bibfnamefont
  {Q.-R.}\ \bibnamefont {Zheng}}, \ and\ \bibinfo {author} {\bibfnamefont
  {G.}~\bibnamefont {Su}},\ }\href@noop {} {\bibfield  {journal} {\bibinfo
  {journal} {arXiv preprint arXiv:1603.01791}\ }\textbf {\bibinfo {volume} {0}}
  (\bibinfo {year} {2016})}\BibitemShut {NoStop}%
\bibitem [{\citenamefont {Kresse}\ and\ \citenamefont
  {Joubert}(1999)}]{PhysRevB.59.1758}%
  \BibitemOpen
  \bibfield  {author} {\bibinfo {author} {\bibfnamefont {G.}~\bibnamefont
  {Kresse}}\ and\ \bibinfo {author} {\bibfnamefont {D.}~\bibnamefont
  {Joubert}},\ }\href {\doibase 10.1103/PhysRevB.59.1758} {\bibfield  {journal}
  {\bibinfo  {journal} {Phys. Rev. B}\ }\textbf {\bibinfo {volume} {59}},\
  \bibinfo {pages} {1758} (\bibinfo {year} {1999})}\BibitemShut {NoStop}%
\bibitem [{\citenamefont {Kresse}\ and\ \citenamefont
  {Furthm\"uller}(1996)}]{PhysRevB.54.11169}%
  \BibitemOpen
  \bibfield  {author} {\bibinfo {author} {\bibfnamefont {G.}~\bibnamefont
  {Kresse}}\ and\ \bibinfo {author} {\bibfnamefont {J.}~\bibnamefont
  {Furthm\"uller}},\ }\href {\doibase 10.1103/PhysRevB.54.11169} {\bibfield
  {journal} {\bibinfo  {journal} {Phys. Rev. B}\ }\textbf {\bibinfo {volume}
  {54}},\ \bibinfo {pages} {11169} (\bibinfo {year} {1996})}\BibitemShut
  {NoStop}%
\bibitem [{\citenamefont {Perdew}\ \emph {et~al.}(1996)\citenamefont {Perdew},
  \citenamefont {Burke},\ and\ \citenamefont
  {Ernzerhof}}]{PhysRevLett.77.3865}%
  \BibitemOpen
  \bibfield  {author} {\bibinfo {author} {\bibfnamefont {J.~P.}\ \bibnamefont
  {Perdew}}, \bibinfo {author} {\bibfnamefont {K.}~\bibnamefont {Burke}}, \
  and\ \bibinfo {author} {\bibfnamefont {M.}~\bibnamefont {Ernzerhof}},\ }\href
  {\doibase 10.1103/PhysRevLett.77.3865} {\bibfield  {journal} {\bibinfo
  {journal} {Phys. Rev. Lett.}\ }\textbf {\bibinfo {volume} {77}},\ \bibinfo
  {pages} {3865} (\bibinfo {year} {1996})}\BibitemShut {NoStop}%
\bibitem [{\citenamefont {Monkhorst}\ and\ \citenamefont
  {Pack}(1976)}]{PhysRevB.13.5188}%
  \BibitemOpen
  \bibfield  {author} {\bibinfo {author} {\bibfnamefont {H.~J.}\ \bibnamefont
  {Monkhorst}}\ and\ \bibinfo {author} {\bibfnamefont {J.~D.}\ \bibnamefont
  {Pack}},\ }\href {\doibase 10.1103/PhysRevB.13.5188} {\bibfield  {journal}
  {\bibinfo  {journal} {Phys. Rev. B}\ }\textbf {\bibinfo {volume} {13}},\
  \bibinfo {pages} {5188} (\bibinfo {year} {1976})}\BibitemShut {NoStop}%
\bibitem [{\citenamefont {Broido}\ \emph {et~al.}(2007)\citenamefont {Broido},
  \citenamefont {Malorny}, \citenamefont {Birner}, \citenamefont {Mingo},\ and\
  \citenamefont {Stewart}}]{ap1.2822891}%
  \BibitemOpen
  \bibfield  {author} {\bibinfo {author} {\bibfnamefont {D.~A.}\ \bibnamefont
  {Broido}}, \bibinfo {author} {\bibfnamefont {M.}~\bibnamefont {Malorny}},
  \bibinfo {author} {\bibfnamefont {G.}~\bibnamefont {Birner}}, \bibinfo
  {author} {\bibfnamefont {N.}~\bibnamefont {Mingo}}, \ and\ \bibinfo {author}
  {\bibfnamefont {D.~A.}\ \bibnamefont {Stewart}},\ }\href {\doibase
  http://dx.doi.org/10.1063/1.2822891} {\bibfield  {journal} {\bibinfo
  {journal} {Appl. Phys. Lett.}\ }\textbf {\bibinfo {volume} {91}},\ \bibinfo
  {eid} {231922} (\bibinfo {year} {2007})}\BibitemShut {NoStop}%
\bibitem [{\citenamefont {Li}\ \emph {et~al.}(2012)\citenamefont {Li},
  \citenamefont {Lindsay}, \citenamefont {Broido}, \citenamefont {Stewart},\
  and\ \citenamefont {Mingo}}]{PhysRevB.86.174307}%
  \BibitemOpen
  \bibfield  {author} {\bibinfo {author} {\bibfnamefont {W.}~\bibnamefont
  {Li}}, \bibinfo {author} {\bibfnamefont {L.}~\bibnamefont {Lindsay}},
  \bibinfo {author} {\bibfnamefont {D.~A.}\ \bibnamefont {Broido}}, \bibinfo
  {author} {\bibfnamefont {D.~A.}\ \bibnamefont {Stewart}}, \ and\ \bibinfo
  {author} {\bibfnamefont {N.}~\bibnamefont {Mingo}},\ }\href {\doibase
  10.1103/PhysRevB.86.174307} {\bibfield  {journal} {\bibinfo  {journal} {Phys.
  Rev. B}\ }\textbf {\bibinfo {volume} {86}},\ \bibinfo {pages} {174307}
  (\bibinfo {year} {2012})}\BibitemShut {NoStop}%
\bibitem [{\citenamefont {Lindsay}\ \emph {et~al.}(2014)\citenamefont
  {Lindsay}, \citenamefont {Li}, \citenamefont {Carrete}, \citenamefont
  {Mingo}, \citenamefont {Broido},\ and\ \citenamefont
  {Reinecke}}]{PhysRevB.89.155426}%
  \BibitemOpen
  \bibfield  {author} {\bibinfo {author} {\bibfnamefont {L.}~\bibnamefont
  {Lindsay}}, \bibinfo {author} {\bibfnamefont {W.}~\bibnamefont {Li}},
  \bibinfo {author} {\bibfnamefont {J.}~\bibnamefont {Carrete}}, \bibinfo
  {author} {\bibfnamefont {N.}~\bibnamefont {Mingo}}, \bibinfo {author}
  {\bibfnamefont {D.~A.}\ \bibnamefont {Broido}}, \ and\ \bibinfo {author}
  {\bibfnamefont {T.~L.}\ \bibnamefont {Reinecke}},\ }\href {\doibase
  10.1103/PhysRevB.89.155426} {\bibfield  {journal} {\bibinfo  {journal} {Phys.
  Rev. B}\ }\textbf {\bibinfo {volume} {89}},\ \bibinfo {pages} {155426}
  (\bibinfo {year} {2014})}\BibitemShut {NoStop}%
\bibitem [{\citenamefont {Li}\ \emph {et~al.}(2014)\citenamefont {Li},
  \citenamefont {Carrete}, \citenamefont {Katcho},\ and\ \citenamefont
  {Mingo}}]{Li20141747}%
  \BibitemOpen
  \bibfield  {author} {\bibinfo {author} {\bibfnamefont {W.}~\bibnamefont
  {Li}}, \bibinfo {author} {\bibfnamefont {J.}~\bibnamefont {Carrete}},
  \bibinfo {author} {\bibfnamefont {N.~A.}\ \bibnamefont {Katcho}}, \ and\
  \bibinfo {author} {\bibfnamefont {N.}~\bibnamefont {Mingo}},\ }\href
  {\doibase http://dx.doi.org/10.1016/j.cpc.2014.02.015} {\bibfield  {journal}
  {\bibinfo  {journal} {Comput. Phys. Commun.}\ }\textbf {\bibinfo {volume}
  {185}},\ \bibinfo {pages} {1747 } (\bibinfo {year} {2014})}\BibitemShut
  {NoStop}%
\bibitem [{\citenamefont {Ward}\ and\ \citenamefont
  {Broido}(2010)}]{PhysRevB.81.085205}%
  \BibitemOpen
  \bibfield  {author} {\bibinfo {author} {\bibfnamefont {A.}~\bibnamefont
  {Ward}}\ and\ \bibinfo {author} {\bibfnamefont {D.~A.}\ \bibnamefont
  {Broido}},\ }\href {\doibase 10.1103/PhysRevB.81.085205} {\bibfield
  {journal} {\bibinfo  {journal} {Phys. Rev. B}\ }\textbf {\bibinfo {volume}
  {81}},\ \bibinfo {pages} {085205} (\bibinfo {year} {2010})}\BibitemShut
  {NoStop}%
\bibitem [{\citenamefont {Ward}\ \emph {et~al.}(2009)\citenamefont {Ward},
  \citenamefont {Broido}, \citenamefont {Stewart},\ and\ \citenamefont
  {Deinzer}}]{PhysRevB.80.125203}%
  \BibitemOpen
  \bibfield  {author} {\bibinfo {author} {\bibfnamefont {A.}~\bibnamefont
  {Ward}}, \bibinfo {author} {\bibfnamefont {D.~A.}\ \bibnamefont {Broido}},
  \bibinfo {author} {\bibfnamefont {D.~A.}\ \bibnamefont {Stewart}}, \ and\
  \bibinfo {author} {\bibfnamefont {G.}~\bibnamefont {Deinzer}},\ }\href
  {\doibase 10.1103/PhysRevB.80.125203} {\bibfield  {journal} {\bibinfo
  {journal} {Phys. Rev. B}\ }\textbf {\bibinfo {volume} {80}},\ \bibinfo
  {pages} {125203} (\bibinfo {year} {2009})}\BibitemShut {NoStop}%
\bibitem [{\citenamefont {Omini}\ and\ \citenamefont
  {Sparavigna}(1996)}]{Phys.Rev.B.1996.53.14.9064-9073}%
  \BibitemOpen
  \bibfield  {author} {\bibinfo {author} {\bibfnamefont {M.}~\bibnamefont
  {Omini}}\ and\ \bibinfo {author} {\bibfnamefont {A.}~\bibnamefont
  {Sparavigna}},\ }\href {\doibase 10.1103/PhysRevB.53.9064} {\bibfield
  {journal} {\bibinfo  {journal} {Phys. Rev. B}\ }\textbf {\bibinfo {volume}
  {53}},\ \bibinfo {pages} {9064} (\bibinfo {year} {1996})}\BibitemShut
  {NoStop}%
\bibitem [{\citenamefont {Lindsay}\ and\ \citenamefont
  {Broido}(2008)}]{0953-8984-20-16-165209}%
  \BibitemOpen
  \bibfield  {author} {\bibinfo {author} {\bibfnamefont {L.}~\bibnamefont
  {Lindsay}}\ and\ \bibinfo {author} {\bibfnamefont {D.~A.}\ \bibnamefont
  {Broido}},\ }\href {http://stacks.iop.org/0953-8984/20/i=16/a=165209}
  {\bibfield  {journal} {\bibinfo  {journal} {Journal of Physics: Condensed
  Matter}\ }\textbf {\bibinfo {volume} {20}},\ \bibinfo {pages} {165209}
  (\bibinfo {year} {2008})}\BibitemShut {NoStop}%
\bibitem [{\citenamefont {Kundu}\ \emph {et~al.}(2011)\citenamefont {Kundu},
  \citenamefont {Mingo}, \citenamefont {Broido},\ and\ \citenamefont
  {Stewart}}]{PhysRevB.84.125426}%
  \BibitemOpen
  \bibfield  {author} {\bibinfo {author} {\bibfnamefont {A.}~\bibnamefont
  {Kundu}}, \bibinfo {author} {\bibfnamefont {N.}~\bibnamefont {Mingo}},
  \bibinfo {author} {\bibfnamefont {D.~A.}\ \bibnamefont {Broido}}, \ and\
  \bibinfo {author} {\bibfnamefont {D.~A.}\ \bibnamefont {Stewart}},\ }\href
  {\doibase 10.1103/PhysRevB.84.125426} {\bibfield  {journal} {\bibinfo
  {journal} {Phys. Rev. B}\ }\textbf {\bibinfo {volume} {84}},\ \bibinfo
  {pages} {125426} (\bibinfo {year} {2011})}\BibitemShut {NoStop}%
\bibitem [{\citenamefont {Tamura}(1983)}]{PhysRevB.27.858}%
  \BibitemOpen
  \bibfield  {author} {\bibinfo {author} {\bibfnamefont {S.-i.}\ \bibnamefont
  {Tamura}},\ }\href {\doibase 10.1103/PhysRevB.27.858} {\bibfield  {journal}
  {\bibinfo  {journal} {Phys. Rev. B}\ }\textbf {\bibinfo {volume} {27}},\
  \bibinfo {pages} {858} (\bibinfo {year} {1983})}\BibitemShut {NoStop}%
\bibitem [{\citenamefont {Qin}\ \emph {et~al.}(2014)\citenamefont {Qin},
  \citenamefont {Yan}, \citenamefont {Qin}, \citenamefont {Yue}, \citenamefont
  {Cui}, \citenamefont {Zheng},\ and\ \citenamefont {Su}}]{QINsrep046946}%
  \BibitemOpen
  \bibfield  {author} {\bibinfo {author} {\bibfnamefont {G.}~\bibnamefont
  {Qin}}, \bibinfo {author} {\bibfnamefont {Q.-B.}\ \bibnamefont {Yan}},
  \bibinfo {author} {\bibfnamefont {Z.}~\bibnamefont {Qin}}, \bibinfo {author}
  {\bibfnamefont {S.-Y.}\ \bibnamefont {Yue}}, \bibinfo {author} {\bibfnamefont
  {H.-J.}\ \bibnamefont {Cui}}, \bibinfo {author} {\bibfnamefont {Q.-R.}\
  \bibnamefont {Zheng}}, \ and\ \bibinfo {author} {\bibfnamefont
  {G.}~\bibnamefont {Su}},\ }\href@noop {} {\bibfield  {journal} {\bibinfo
  {journal} {Sci. Rep.}\ }\textbf {\bibinfo {volume} {4}},\ \bibinfo {eid}
  {6946} (\bibinfo {year} {2014})}\BibitemShut {NoStop}%
\bibitem [{\citenamefont {Qin}\ \emph {et~al.}(2015)\citenamefont {Qin},
  \citenamefont {Yan}, \citenamefont {Qin}, \citenamefont {Yue}, \citenamefont
  {Hu},\ and\ \citenamefont {Su}}]{Phys.Chem.Chem.Phys..2015..17.4854}%
  \BibitemOpen
  \bibfield  {author} {\bibinfo {author} {\bibfnamefont {G.}~\bibnamefont
  {Qin}}, \bibinfo {author} {\bibfnamefont {Q.-B.}\ \bibnamefont {Yan}},
  \bibinfo {author} {\bibfnamefont {Z.}~\bibnamefont {Qin}}, \bibinfo {author}
  {\bibfnamefont {S.-Y.}\ \bibnamefont {Yue}}, \bibinfo {author} {\bibfnamefont
  {M.}~\bibnamefont {Hu}}, \ and\ \bibinfo {author} {\bibfnamefont
  {G.}~\bibnamefont {Su}},\ }\href {\doibase 10.1039/C4CP04858J} {\bibfield
  {journal} {\bibinfo  {journal} {Phys. Chem. Chem. Phys.}\ }\textbf {\bibinfo
  {volume} {17}},\ \bibinfo {pages} {4854} (\bibinfo {year}
  {2015})}\BibitemShut {NoStop}%
\bibitem [{\citenamefont {Yin}\ and\ \citenamefont
  {Cohen}(1982)}]{PhysRevB.26.3259}%
  \BibitemOpen
  \bibfield  {author} {\bibinfo {author} {\bibfnamefont {M.~T.}\ \bibnamefont
  {Yin}}\ and\ \bibinfo {author} {\bibfnamefont {M.~L.}\ \bibnamefont
  {Cohen}},\ }\href {\doibase 10.1103/PhysRevB.26.3259} {\bibfield  {journal}
  {\bibinfo  {journal} {Phys. Rev. B}\ }\textbf {\bibinfo {volume} {26}},\
  \bibinfo {pages} {3259} (\bibinfo {year} {1982})}\BibitemShut {NoStop}%
\bibitem [{\citenamefont {Togo}\ \emph {et~al.}(2008)\citenamefont {Togo},
  \citenamefont {Oba},\ and\ \citenamefont {Tanaka}}]{phonopy}%
  \BibitemOpen
  \bibfield  {author} {\bibinfo {author} {\bibfnamefont {A.}~\bibnamefont
  {Togo}}, \bibinfo {author} {\bibfnamefont {F.}~\bibnamefont {Oba}}, \ and\
  \bibinfo {author} {\bibfnamefont {I.}~\bibnamefont {Tanaka}},\ }\href@noop {}
  {\bibfield  {journal} {\bibinfo  {journal} {Phys. Rev. B}\ }\textbf {\bibinfo
  {volume} {78}},\ \bibinfo {pages} {134106} (\bibinfo {year}
  {2008})}\BibitemShut {NoStop}%
\bibitem [{\citenamefont {Lindsay}\ \emph {et~al.}(2010)\citenamefont
  {Lindsay}, \citenamefont {Broido},\ and\ \citenamefont
  {Mingo}}]{PhysRevB.82.115427}%
  \BibitemOpen
  \bibfield  {author} {\bibinfo {author} {\bibfnamefont {L.}~\bibnamefont
  {Lindsay}}, \bibinfo {author} {\bibfnamefont {D.~A.}\ \bibnamefont {Broido}},
  \ and\ \bibinfo {author} {\bibfnamefont {N.}~\bibnamefont {Mingo}},\ }\href
  {\doibase 10.1103/PhysRevB.82.115427} {\bibfield  {journal} {\bibinfo
  {journal} {Phys. Rev. B}\ }\textbf {\bibinfo {volume} {82}},\ \bibinfo
  {pages} {115427} (\bibinfo {year} {2010})}\BibitemShut {NoStop}%
\bibitem [{\citenamefont {Castro~Neto}\ \emph {et~al.}(2009)\citenamefont
  {Castro~Neto}, \citenamefont {Guinea}, \citenamefont {Peres}, \citenamefont
  {Novoselov},\ and\ \citenamefont {Geim}}]{RevModPhys.81.109}%
  \BibitemOpen
  \bibfield  {author} {\bibinfo {author} {\bibfnamefont {A.~H.}\ \bibnamefont
  {Castro~Neto}}, \bibinfo {author} {\bibfnamefont {F.}~\bibnamefont {Guinea}},
  \bibinfo {author} {\bibfnamefont {N.~M.~R.}\ \bibnamefont {Peres}}, \bibinfo
  {author} {\bibfnamefont {K.~S.}\ \bibnamefont {Novoselov}}, \ and\ \bibinfo
  {author} {\bibfnamefont {A.~K.}\ \bibnamefont {Geim}},\ }\href {\doibase
  10.1103/RevModPhys.81.109} {\bibfield  {journal} {\bibinfo  {journal} {Rev.
  Mod. Phys.}\ }\textbf {\bibinfo {volume} {81}},\ \bibinfo {pages} {109}
  (\bibinfo {year} {2009})}\BibitemShut {NoStop}%
\bibitem [{\citenamefont {Zhang}\ \emph {et~al.}(2014)\citenamefont {Zhang},
  \citenamefont {Xie}, \citenamefont {Hu}, \citenamefont {Bao}, \citenamefont
  {Yue}, \citenamefont {Qin},\ and\ \citenamefont {Su}}]{PhysRevB.89.054310}%
  \BibitemOpen
  \bibfield  {author} {\bibinfo {author} {\bibfnamefont {X.}~\bibnamefont
  {Zhang}}, \bibinfo {author} {\bibfnamefont {H.}~\bibnamefont {Xie}}, \bibinfo
  {author} {\bibfnamefont {M.}~\bibnamefont {Hu}}, \bibinfo {author}
  {\bibfnamefont {H.}~\bibnamefont {Bao}}, \bibinfo {author} {\bibfnamefont
  {S.}~\bibnamefont {Yue}}, \bibinfo {author} {\bibfnamefont {G.}~\bibnamefont
  {Qin}}, \ and\ \bibinfo {author} {\bibfnamefont {G.}~\bibnamefont {Su}},\
  }\href {\doibase 10.1103/PhysRevB.89.054310} {\bibfield  {journal} {\bibinfo
  {journal} {Phys. Rev. B}\ }\textbf {\bibinfo {volume} {89}},\ \bibinfo
  {pages} {054310} (\bibinfo {year} {2014})}\BibitemShut {NoStop}%
\bibitem [{\citenamefont {Xie}\ \emph {et~al.}(2014)\citenamefont {Xie},
  \citenamefont {Hu},\ and\ \citenamefont
  {Bao}}]{AppliedPhysicsLetters.2014.104.13.131906}%
  \BibitemOpen
  \bibfield  {author} {\bibinfo {author} {\bibfnamefont {H.}~\bibnamefont
  {Xie}}, \bibinfo {author} {\bibfnamefont {M.}~\bibnamefont {Hu}}, \ and\
  \bibinfo {author} {\bibfnamefont {H.}~\bibnamefont {Bao}},\ }\href {\doibase
  http://dx.doi.org/10.1063/1.4870586} {\bibfield  {journal} {\bibinfo
  {journal} {Appl. Phys. Lett.}\ }\textbf {\bibinfo {volume} {104}},\ \bibinfo
  {pages} {131906} (\bibinfo {year} {2014})}\BibitemShut {NoStop}%
\bibitem [{\citenamefont {Xie}\ \emph {et~al.}(2016)\citenamefont {Xie},
  \citenamefont {Ouyang}, \citenamefont {Germaneau}, \citenamefont {Qin},
  \citenamefont {Hu},\ and\ \citenamefont {Bao}}]{Phys.Rev.B.2016.93.7.075404}%
  \BibitemOpen
  \bibfield  {author} {\bibinfo {author} {\bibfnamefont {H.}~\bibnamefont
  {Xie}}, \bibinfo {author} {\bibfnamefont {T.}~\bibnamefont {Ouyang}},
  \bibinfo {author} {\bibfnamefont {?.}~\bibnamefont {Germaneau}}, \bibinfo
  {author} {\bibfnamefont {G.}~\bibnamefont {Qin}}, \bibinfo {author}
  {\bibfnamefont {M.}~\bibnamefont {Hu}}, \ and\ \bibinfo {author}
  {\bibfnamefont {H.}~\bibnamefont {Bao}},\ }\href {\doibase
  10.1103/PhysRevB.93.075404} {\bibfield  {journal} {\bibinfo  {journal} {Phys.
  Rev. B}\ }\textbf {\bibinfo {volume} {93}},\ \bibinfo {pages} {075404}
  (\bibinfo {year} {2016})}\BibitemShut {NoStop}%
\bibitem [{\citenamefont {Parker}\ and\ \citenamefont
  {Singh}(2010)}]{PhysRevB.82.035204}%
  \BibitemOpen
  \bibfield  {author} {\bibinfo {author} {\bibfnamefont {D.}~\bibnamefont
  {Parker}}\ and\ \bibinfo {author} {\bibfnamefont {D.~J.}\ \bibnamefont
  {Singh}},\ }\href {\doibase 10.1103/PhysRevB.82.035204} {\bibfield  {journal}
  {\bibinfo  {journal} {Phys. Rev. B}\ }\textbf {\bibinfo {volume} {82}},\
  \bibinfo {pages} {035204} (\bibinfo {year} {2010})}\BibitemShut {NoStop}%
\bibitem [{\citenamefont {Pulikkotil}\ \emph {et~al.}(2012)\citenamefont
  {Pulikkotil}, \citenamefont {Singh}, \citenamefont {Auluck}, \citenamefont
  {Saravanan}, \citenamefont {Misra}, \citenamefont {Dhar},\ and\ \citenamefont
  {Budhani}}]{PhysRevB.86.155204}%
  \BibitemOpen
  \bibfield  {author} {\bibinfo {author} {\bibfnamefont {J.~J.}\ \bibnamefont
  {Pulikkotil}}, \bibinfo {author} {\bibfnamefont {D.~J.}\ \bibnamefont
  {Singh}}, \bibinfo {author} {\bibfnamefont {S.}~\bibnamefont {Auluck}},
  \bibinfo {author} {\bibfnamefont {M.}~\bibnamefont {Saravanan}}, \bibinfo
  {author} {\bibfnamefont {D.~K.}\ \bibnamefont {Misra}}, \bibinfo {author}
  {\bibfnamefont {A.}~\bibnamefont {Dhar}}, \ and\ \bibinfo {author}
  {\bibfnamefont {R.~C.}\ \bibnamefont {Budhani}},\ }\href {\doibase
  10.1103/PhysRevB.86.155204} {\bibfield  {journal} {\bibinfo  {journal} {Phys.
  Rev. B}\ }\textbf {\bibinfo {volume} {86}},\ \bibinfo {pages} {155204}
  (\bibinfo {year} {2012})}\BibitemShut {NoStop}%
\bibitem [{\citenamefont {Li}\ \emph {et~al.}(2013{\natexlab{b}})\citenamefont
  {Li}, \citenamefont {Carrete},\ and\ \citenamefont
  {Mingo}}]{AppliedPhysicsLetters.2013.103.25.253103}%
  \BibitemOpen
  \bibfield  {author} {\bibinfo {author} {\bibfnamefont {W.}~\bibnamefont
  {Li}}, \bibinfo {author} {\bibfnamefont {J.}~\bibnamefont {Carrete}}, \ and\
  \bibinfo {author} {\bibfnamefont {N.}~\bibnamefont {Mingo}},\ }\href
  {\doibase http://dx.doi.org/10.1063/1.4850995} {\bibfield  {journal}
  {\bibinfo  {journal} {Appl. Phys. Lett.}\ }\textbf {\bibinfo {volume}
  {103}},\ \bibinfo {pages} {253103} (\bibinfo {year}
  {2013}{\natexlab{b}})}\BibitemShut {NoStop}%
\bibitem [{\citenamefont {Esfarjani}\ \emph {et~al.}(2011)\citenamefont
  {Esfarjani}, \citenamefont {Chen},\ and\ \citenamefont
  {Stokes}}]{PhysRevB.84.085204}%
  \BibitemOpen
  \bibfield  {author} {\bibinfo {author} {\bibfnamefont {K.}~\bibnamefont
  {Esfarjani}}, \bibinfo {author} {\bibfnamefont {G.}~\bibnamefont {Chen}}, \
  and\ \bibinfo {author} {\bibfnamefont {H.~T.}\ \bibnamefont {Stokes}},\
  }\href {\doibase 10.1103/PhysRevB.84.085204} {\bibfield  {journal} {\bibinfo
  {journal} {Phys. Rev. B}\ }\textbf {\bibinfo {volume} {84}},\ \bibinfo
  {pages} {085204} (\bibinfo {year} {2011})}\BibitemShut {NoStop}%
\bibitem [{\citenamefont {Lindsay}\ \emph {et~al.}(2013)\citenamefont
  {Lindsay}, \citenamefont {Broido},\ and\ \citenamefont
  {Reinecke}}]{PhysRevLett.111.025901}%
  \BibitemOpen
  \bibfield  {author} {\bibinfo {author} {\bibfnamefont {L.}~\bibnamefont
  {Lindsay}}, \bibinfo {author} {\bibfnamefont {D.~A.}\ \bibnamefont {Broido}},
  \ and\ \bibinfo {author} {\bibfnamefont {T.~L.}\ \bibnamefont {Reinecke}},\
  }\href {\doibase 10.1103/PhysRevLett.111.025901} {\bibfield  {journal}
  {\bibinfo  {journal} {Phys. Rev. Lett.}\ }\textbf {\bibinfo {volume} {111}},\
  \bibinfo {pages} {025901} (\bibinfo {year} {2013})}\BibitemShut {NoStop}%
\bibitem [{\citenamefont {Li}\ \emph {et~al.}(2015)\citenamefont {Li},
  \citenamefont {Hong}, \citenamefont {May}, \citenamefont {Bansal},
  \citenamefont {Chi}, \citenamefont {Hong}, \citenamefont {Ehlers},\ and\
  \citenamefont {Delaire}}]{NatPhys.2015.11.12.1063-1069}%
  \BibitemOpen
  \bibfield  {author} {\bibinfo {author} {\bibfnamefont {C.~W.}\ \bibnamefont
  {Li}}, \bibinfo {author} {\bibfnamefont {J.}~\bibnamefont {Hong}}, \bibinfo
  {author} {\bibfnamefont {A.~F.}\ \bibnamefont {May}}, \bibinfo {author}
  {\bibfnamefont {D.}~\bibnamefont {Bansal}}, \bibinfo {author} {\bibfnamefont
  {S.}~\bibnamefont {Chi}}, \bibinfo {author} {\bibfnamefont {T.}~\bibnamefont
  {Hong}}, \bibinfo {author} {\bibfnamefont {G.}~\bibnamefont {Ehlers}}, \ and\
  \bibinfo {author} {\bibfnamefont {O.}~\bibnamefont {Delaire}},\ }\href
  {\doibase 10.1038/nphys3492} {\bibfield  {journal} {\bibinfo  {journal} {Nat
  Phys}\ }\textbf {\bibinfo {volume} {11}},\ \bibinfo {pages} {1063} (\bibinfo
  {year} {2015})}\BibitemShut {NoStop}%
\bibitem [{\citenamefont {Guo}\ \emph {et~al.}(2015{\natexlab{b}})\citenamefont
  {Guo}, \citenamefont {Verma}, \citenamefont {Wu}, \citenamefont {Sun},
  \citenamefont {Hickman}, \citenamefont {Masui}, \citenamefont {Kuramata},
  \citenamefont {Higashiwaki}, \citenamefont {Jena},\ and\ \citenamefont
  {Luo}}]{AppliedPhysicsLetters.2015.106.11.111909}%
  \BibitemOpen
  \bibfield  {author} {\bibinfo {author} {\bibfnamefont {Z.}~\bibnamefont
  {Guo}}, \bibinfo {author} {\bibfnamefont {A.}~\bibnamefont {Verma}}, \bibinfo
  {author} {\bibfnamefont {X.}~\bibnamefont {Wu}}, \bibinfo {author}
  {\bibfnamefont {F.}~\bibnamefont {Sun}}, \bibinfo {author} {\bibfnamefont
  {A.}~\bibnamefont {Hickman}}, \bibinfo {author} {\bibfnamefont
  {T.}~\bibnamefont {Masui}}, \bibinfo {author} {\bibfnamefont
  {A.}~\bibnamefont {Kuramata}}, \bibinfo {author} {\bibfnamefont
  {M.}~\bibnamefont {Higashiwaki}}, \bibinfo {author} {\bibfnamefont
  {D.}~\bibnamefont {Jena}}, \ and\ \bibinfo {author} {\bibfnamefont
  {T.}~\bibnamefont {Luo}},\ }\href {\doibase
  http://dx.doi.org/10.1063/1.4916078} {\bibfield  {journal} {\bibinfo
  {journal} {Appl. Phys. Lett.}\ }\textbf {\bibinfo {volume} {106}},\ \bibinfo
  {pages} {111909} (\bibinfo {year} {2015}{\natexlab{b}})}\BibitemShut
  {NoStop}%
\bibitem [{\citenamefont {Savin}\ \emph {et~al.}(1997)\citenamefont {Savin},
  \citenamefont {Nesper}, \citenamefont {Wengert},\ and\ \citenamefont
  {Fässler}}]{Angew.Chem.Int.Ed.Engl..1997.36.17.1808-1832}%
  \BibitemOpen
  \bibfield  {author} {\bibinfo {author} {\bibfnamefont {A.}~\bibnamefont
  {Savin}}, \bibinfo {author} {\bibfnamefont {R.}~\bibnamefont {Nesper}},
  \bibinfo {author} {\bibfnamefont {S.}~\bibnamefont {Wengert}}, \ and\
  \bibinfo {author} {\bibfnamefont {T.~F.}\ \bibnamefont {Fässler}},\ }\href
  {\doibase 10.1002/anie.199718081} {\bibfield  {journal} {\bibinfo  {journal}
  {Angew. Chem. Int. Ed. Engl.}\ }\textbf {\bibinfo {volume} {36}},\ \bibinfo
  {pages} {1808} (\bibinfo {year} {1997})}\BibitemShut {NoStop}%
\bibitem [{\citenamefont {Steinmann}\ \emph {et~al.}(2011)\citenamefont
  {Steinmann}, \citenamefont {Mo},\ and\ \citenamefont
  {Corminboeuf}}]{C1CP21055F}%
  \BibitemOpen
  \bibfield  {author} {\bibinfo {author} {\bibfnamefont {S.~N.}\ \bibnamefont
  {Steinmann}}, \bibinfo {author} {\bibfnamefont {Y.}~\bibnamefont {Mo}}, \
  and\ \bibinfo {author} {\bibfnamefont {C.}~\bibnamefont {Corminboeuf}},\
  }\href {\doibase 10.1039/C1CP21055F} {\bibfield  {journal} {\bibinfo
  {journal} {Phys. Chem. Chem. Phys.}\ }\textbf {\bibinfo {volume} {13}},\
  \bibinfo {pages} {20584} (\bibinfo {year} {2011})}\BibitemShut {NoStop}%
\bibitem [{\citenamefont {Ziman}(1961)}]{ziman1961electrons}%
  \BibitemOpen
  \bibfield  {author} {\bibinfo {author} {\bibfnamefont {J.}~\bibnamefont
  {Ziman}},\ }\href@noop {} {\emph {\bibinfo {title} {Electrons and phonons}}}\
  (\bibinfo  {publisher} {Oxford University Press, London},\ \bibinfo {year}
  {1961})\BibitemShut {NoStop}%
\bibitem [{\citenamefont {Srivastava}(1990)}]{srivastava1990physics}%
  \BibitemOpen
  \bibfield  {author} {\bibinfo {author} {\bibfnamefont {G.~P.}\ \bibnamefont
  {Srivastava}},\ }\href@noop {} {\emph {\bibinfo {title} {The physics of
  phonons}}}\ (\bibinfo  {publisher} {CRC Press},\ \bibinfo {year}
  {1990})\BibitemShut {NoStop}%
\bibitem [{\citenamefont {Nika}\ \emph {et~al.}(2009)\citenamefont {Nika},
  \citenamefont {Pokatilov}, \citenamefont {Askerov},\ and\ \citenamefont
  {Balandin}}]{PhysRevB.79.155413}%
  \BibitemOpen
  \bibfield  {author} {\bibinfo {author} {\bibfnamefont {D.~L.}\ \bibnamefont
  {Nika}}, \bibinfo {author} {\bibfnamefont {E.~P.}\ \bibnamefont {Pokatilov}},
  \bibinfo {author} {\bibfnamefont {A.~S.}\ \bibnamefont {Askerov}}, \ and\
  \bibinfo {author} {\bibfnamefont {A.~A.}\ \bibnamefont {Balandin}},\ }\href
  {\doibase 10.1103/PhysRevB.79.155413} {\bibfield  {journal} {\bibinfo
  {journal} {Phys. Rev. B}\ }\textbf {\bibinfo {volume} {79}},\ \bibinfo
  {pages} {155413} (\bibinfo {year} {2009})}\BibitemShut {NoStop}%
\bibitem [{\citenamefont {Lindsay}\ \emph {et~al.}(2011)\citenamefont
  {Lindsay}, \citenamefont {Broido},\ and\ \citenamefont
  {Mingo}}]{PhysRevB.83.235428}%
  \BibitemOpen
  \bibfield  {author} {\bibinfo {author} {\bibfnamefont {L.}~\bibnamefont
  {Lindsay}}, \bibinfo {author} {\bibfnamefont {D.~A.}\ \bibnamefont {Broido}},
  \ and\ \bibinfo {author} {\bibfnamefont {N.}~\bibnamefont {Mingo}},\ }\href
  {\doibase 10.1103/PhysRevB.83.235428} {\bibfield  {journal} {\bibinfo
  {journal} {Phys. Rev. B}\ }\textbf {\bibinfo {volume} {83}},\ \bibinfo
  {pages} {235428} (\bibinfo {year} {2011})}\BibitemShut {NoStop}%
\bibitem [{\citenamefont {Dong}\ \emph {et~al.}(2015)\citenamefont {Dong},
  \citenamefont {Cao},\ and\ \citenamefont {Guo}}]{PhysicaE.2015.66..1-6}%
  \BibitemOpen
  \bibfield  {author} {\bibinfo {author} {\bibfnamefont {Y.}~\bibnamefont
  {Dong}}, \bibinfo {author} {\bibfnamefont {B.-Y.}\ \bibnamefont {Cao}}, \
  and\ \bibinfo {author} {\bibfnamefont {Z.-Y.}\ \bibnamefont {Guo}},\ }\href
  {\doibase http://dx.doi.org/10.1016/j.physe.2014.09.011} {\bibfield
  {journal} {\bibinfo  {journal} {Physica E}\ }\textbf {\bibinfo {volume}
  {66}},\ \bibinfo {pages} {1} (\bibinfo {year} {2015})}\BibitemShut {NoStop}%
\end{thebibliography}

%

\end{document}